\documentclass[12pt]{article}
\usepackage{epsf}
\usepackage{graphicx}
\usepackage{amssymb}
\def\uu{\langle \bar u u \rangle}
\def\dd{\langle \bar d d \rangle}
\def\ss{\langle \bar s s \rangle}
\def\qq{\langle \bar q q \rangle}

\def\es{\!\!\! &=& \!\!\!}

\def\nnb{\nonumber}

\title{ Radiative $\Omega_{Q}^{*}\rightarrow\Omega_{Q}\gamma$ and
$\Xi_{Q}^{*}\rightarrow\Xi^{\prime}_{Q}\gamma$ transitions in  light cone QCD}

\author{
  T. M. Aliev\footnote{Permanent address: Institute of Physics, Baku,
 Azerbaijan} \thanks{ e-mail: taliev@metu.edu.tr},
  K. Azizi\thanks {e-mail:kazizi@dogus.edu.tr, kazem.azizi@cern.ch},
  H. Sundu \thanks {e-mail:hayriye.sundu@kocaeli.edu.tr} \\
  \small$^\dag$ Department of Physics, Middle East Technical University, 06531, Ankara,
Turkey \\
  \small$\ddag$ Department of Physics, Do\u gu\c s University, Ac{\i}badem-Kad{\i}k\"oy, 34722
  Istanbul, Turkey \\
  \small $\S$ Department of Physics, Kocaeli University, 41380
Izmit, Turkey}
 \date{}

\begin{document}
\setlength{\baselineskip}{26pt} \maketitle
\setlength{\baselineskip}{7mm}
\begin{abstract}

\end{abstract}
We calculate the magnetic dipole and electric quadrupole moments associated with the radiative  $\Omega_{Q}^{*}\rightarrow\Omega_{Q}\gamma$ and
$\Xi_{Q}^{*}\rightarrow\Xi^{\prime}_{Q}\gamma$  transitions with $Q=b$ or $c$ in  the framework of light cone QCD sum rules. It is found that the corresponding  quadrupole moments are negligibly small while the
magnetic dipole moments are considerably large.  A comparison of the results on the considered multipole moments as well as corresponding decay widths with the predictions of the vector dominance model is performed.

 \thispagestyle{empty}
\newpage
\setcounter{page}{1}
\section{Introduction}
In the recent years, there have been significant experimental progresses on hadron spectroscopy.  Many new baryons containing heavy bottom and charm quarks as well as many new charmonium like states are observed. Now, all heavy baryons with single
heavy quark have been discovered in the experiments except the $\Omega^*_b$ baryon with spin--3/2. In the case of doubly heavy baryons only the doubly charmed  $\Xi_{cc}$ baryon has been discovered
 by SELEX Collaboration \cite{Rbwtq02,Rbwtq03} but the experimental attempts on the identification of other members of the doubly baryons as well as  triply heavy baryons predicted by quark model
are continued. Considering these progresses and facilities of experiments specially at LHC, it would be possible to study the decay properties of heavy baryons  in near future.
Theoretical studies on  electromagnetic, weak and strong decays of heavy baryons  receive special attention in the light of the experimental results.

  In the present work we calculate the electromagnetic form factors of the radiative $\Omega_{Q}^{*}\rightarrow\Omega_{Q}\gamma$ and
$\Xi_{Q}^{*}\rightarrow\Xi^{\prime}_{Q}\gamma$ transitions in the framework of the  light cone QCD sum rules as one of the most applicable non-perturbative tools to hadron physics. Here, baryons with $*$ correspond
spin--3/2 while those without $*$ are spin--1/2 baryons. Using the electromagnetic form factors
at static limit ($q^2=0$), we obtain the magnetic dipole and electric quadrupole moments as well as decay widths of the considered radiative decays. We compare our results with the predictions of the vector meson dominance model (VDM)
\cite{zamir} which  uses  the values of the strong coupling constants  between spin--3/2 and  spin--1/2 heavy baryons with vector mesons \cite{zamir2} to calculate the magnetic dipole and electric quadrupole moments
 of the transitions under consideration. The electromagnetic multipole moments of heavy baryons can give valuable information
 on their internal structure as well as their geometric shapes. Note that  other possible  radiative transitions among heavy spin--3/2 and spin--1/2 baryons with single heavy quark, namely
$\Sigma_{Q}^{*}\rightarrow \Sigma_{Q}\gamma$ , $\Xi_{Q}^{*}\rightarrow \Xi_{Q}\gamma$  and $\Sigma_{Q}^{*}\rightarrow\Lambda_{Q}\gamma$
have been investigated in  \cite{azizi} using the same framework.  Some of these radiative transitions have also been previously studied using chiral perturbation theory \cite{banuls}, heavy
quark and chiral symmetries \cite{cheng,tawfig},  relativistic quark model \cite{ivanov} and
light cone QCD sum rules at leading order in HQET in \cite{zhu}.

The outline of the paper is as follows. In next section,   QCD sum rules for the electromagnetic form factors
of the transitions under consideration are calculated. In last section, we numerically analyze the obtained  sum rules. This section also includes
comparison of our results with the predictions of VDM on the multipole moments as well as the corresponding decay widths.

\section{Theoretical framework }

The aim of this section is to obtain light cone QCD sum rules (LCQSR) for the electromagnetic form factors defining the radiative  $\Omega_{Q}^{*}\rightarrow\Omega_{Q}\gamma$ and
$\Xi_{Q}^{*}\rightarrow\Xi^{\prime}_{Q}\gamma$ transitions. For this goal  we use the following  two-point correlation
function in the presence of an external photon field:
\begin{equation}\label{T}
\Pi_{\mu}(p,q)=i\int d^{4}xe^{ip\cdot x}\langle0\mid
T\{\eta(x)\bar{\eta}_{\mu}(0) \}\mid0\rangle_{\gamma},
\end{equation}
where $\eta$ and $\eta_{\mu}$ are  the interpolating
currents of the heavy flavored baryons with spin 1/2 and 3/2,
respectively. The main task in the following is to calculate this correlation function once in terms of hadronic parameters called the hadronic side and  in terms
of photon distribution amplitudes (DAs) with increasing twist by the help of operator product expansion (OPE). By equating the coefficients of appropriate structures from hadronic to OPE side, we obtain LCQSR for the
transition form factors. To suppress the contribution of the higher states and continuum, we apply Borel transformations with respect to the momentum squared of the initial and final baryonic states.
For further pushing down those contributions, we also apply continuum subtraction to both sides of the LCQSRs obtained.

\subsection{Hadronic side}

To obtain the hadronic representation,    we insert  complete sets of intermediate states having the same quantum
numbers as the interpolating currents into the above correlation function. As a result of which we get
\begin{eqnarray}\label{T2}
\Pi_{\mu}(p,q)&=&\frac{\langle0\mid \eta\mid
2(p,s')\rangle}{p^{2}-m_{2}^{2}}\langle 2(p,s')\mid
1(p+q,s)\rangle_\gamma\frac{\langle 1(p_+q,s)\mid \bar{\eta}_{\mu}\mid
0\rangle}{(p+q)^{2}-m_{1}^{2}}+...,
\end{eqnarray}
where the dots indicate the contributions of the higher states and
continuum and $q$ is the photon's momentum. In the above equation,
$\langle1(p+q,s)|$ and $\langle2(p,s')|$ denote the heavy spin 3/2 and 1/2
states and $m_{1}$ and $m_{2}$ are their masses, respectively. To proceed,
we need to know the matrix elements of the interpolating currents
between the  vacuum and the baryonic states. They are defined in terms of spinors and residues as
\begin{eqnarray}\label{lambdabey}
\langle1(p+q,s)\mid \bar \eta_{\mu}(0)\mid 0\rangle&=&\lambda_{1}\bar u_{\mu}(p+q,s),
\nonumber\\
\langle 0 \mid \eta (0)\mid 2(p,s')\rangle &=&\lambda_{2} u(p,s'),
\end{eqnarray}
where $u_{\mu}(p,s)$ is the Rarita-Schwinger spinor; and  $\lambda_{1}$
and  $\lambda_{2}$ are the   residues of the heavy baryons with spin 3/2 and 1/2,
respectively which are calculated in \cite{azizi}.  The matrix element $\langle 2(p,s')\mid
1(p+q,s)\rangle_\gamma$ is also defined as
\cite{onalti,onyedi}
\begin{eqnarray}\label{matelpar}
\langle 2(p,s')\mid
1(p+q,s)\rangle_\gamma&=&e\bar u(p,s')\left\{\vphantom{\int_0^{x_2}}G_{1}(q_{\mu}
\not\!\varepsilon-\varepsilon_{\mu}\not\!q)\right.
+G_{2}[({\cal P}\varepsilon)q_{\mu}-({\cal P}q)\varepsilon_{\mu}]\gamma_{5}
\nonumber\\&+&G_{3}[(q\varepsilon)q_{\mu}-q^{2}\varepsilon_{\mu}]\gamma_{5}\left.
\vphantom{\int_0^{x_2}}\right\} u_{\mu}(p+q,s),\nonumber\\
\end{eqnarray}
where $G_i$ are electromagnetic form factors,  $\varepsilon_{\mu}$ is the photon's polarization vector and
${\cal P}=\frac{p+(p+q)}{2}$. In the above equation, the term proportional to $G_{3}$ is zero for  the real photon which we consider in the present study.
 At  $q^2=0$, the  transition magnetic dipole moment
$G_{M}$ and the  electric quadrupole moment $G_{E}$ are defined in terms of the remaining electromagnetic form factors as
\begin{eqnarray}\label{acayip}
 G_{M}&=&\left[(3m_{1}+m_{2})\frac{G_{1}}{m_{1}}+(m_{1}-m_{2})G_{2}\right]
 \frac{m_{2}}{3},\nonumber\\
G_{E}&=&(m_{1}-m_{2})\left[\frac{G_{1}}{m_{1}}+G_{2}\right]\frac{m_{2}}{3}.
\end{eqnarray}

Now, we use Eqs. (4) and (3) in Eq. (2) and perform summation over spins of the Dirac and Rarita-Schwinger spinors.  In the case of spin 3/2 this summation is written as
\begin{equation}\label{raritabela}
\sum_{s}u_{\mu}(p,s)\bar u_{\nu}(p,s)=\frac{(\not\!p+m_)}{2m}\{-g_{\mu\nu}
+\frac{1}{3}\gamma_{\mu}\gamma_{\nu}-\frac{2p_{\mu}p_{\nu}}
{3m^{2}}-\frac{p_{\mu}\gamma_{\nu}-p_{\nu}\gamma_{\mu}}{3m}\}.
\end{equation}
Using Eqs. (3-6), in principle, one can straightforwardly calculate the hadronic side of the correlation function. But, here appear two unwanted problems:
\begin{itemize}
    \item there is pollution from spin--1/2 baryons, since the interpolating current $\eta_{\mu}$ couples with spin--1/2 baryons also.
    \item  All Lorentz structures are not  independent.
\end{itemize}
In order to solve the first problem, let write the corresponding matrix element of the current  $\eta_{\mu}$ between vacuum and $J=1/2$ states, which can be parametrized as
\begin{equation}
 \langle0|\eta_{\mu}|1(p+q,s)\rangle=[\alpha \gamma_{\mu}+\beta (p+q)_{\mu}]u(p+q,s).
\end{equation}
Multiplying both sides of this equation by $\gamma^{\mu}$ and using $\gamma^{\mu}\eta_{\mu}=0$ as well as the Dirac equation we get
\begin{equation}
 \langle0|\eta_{\mu}|1(p+q,s)\rangle=\alpha [\gamma_{\mu}-\frac{4}{m_2} (p+q)_{\mu}]u(p+q,s).
\end{equation}
from this expression it follows that contributions of spin--1/2 states are either proportional to the $\gamma_{\mu}$ at the end or $(p+q)_{\mu}$.
Taking into account this fact, from Eq. (6) it follows that only terms proportional to $g_{\mu\nu}$ contain contributions coming only from spin--3/2 states. This observation shows that how spin--1/2 states'
contributions coupled to $\eta_{\mu}$ can be removed. The second problem can be solved if one orders the Dirac matrices in an appropriate way. In this work,
we choose the ordering  $\not\!\varepsilon\not\!q\not\!p\gamma_{\mu}$. After some calculations, for  the  hadronic side of the correlation function, we get
\begin{eqnarray}\label{final phenpart}
\Pi_\mu &=& e \lambda_{1} \lambda_{2} \frac{1}{p^2-m_{2}^2}
\frac{1}{(p+q)^2-m_{1}^2} \left[ \vphantom{\int_0^{x_2}}\right.
\nonumber \\ && \Big[ \varepsilon_\mu (pq) - (\varepsilon p)
q_\mu \Big] \Big\{ -2 G_1 m_{1} - G_2 m_{1} m_{2} + G_2 (p+q)^2
 \nonumber \\ && +  \Big[ 2 G_1 - G_2(m_{1}-m_{2})
\Big] \not\!p + m_{2} G_2 \not\!q - G_2 \not\!q\not\!p \Big\}
\gamma_5 \nonumber \\ && + \Big[ q_\mu \not\!\varepsilon -
\varepsilon_\mu \not\!q \Big] \Big\{ G_1 (p^2 + m_{1} m_{2}) -
G_1 (m_{1}+m_{2}) \not\!p \Big\} \gamma_5 \nonumber \\ && + 2
G_1 \Big[ \not\!\varepsilon (pq) - \not\!q (\varepsilon p)
\Big] q_\mu \gamma_5  - G_1 \not\!\varepsilon
\not\!q ( m_{2} + \not\!p ) q_\mu \gamma_5 \nonumber
\\ &&
\left.+ \mbox{other structures with $\gamma_\mu$ at the end or which are proportional to $(p+q)_\mu$} \vphantom{\int_0^{x_2}}\right],\nonumber\\
\end{eqnarray}
where, we need two  invariant structures to calculate the form factors
$G_{1}$ and $G_{2}$. In the present work, we select the structures
$\not\!\varepsilon \not\!p\gamma_{5}q_{\mu}$ and
$\not\!q\not\!p\gamma_{5}(\varepsilon p)q_{\mu}$ for $G_{1}$ and $G_{2}$, respectively.
The advantage of these structures is that these terms  do not receive
 contributions from contact terms.

\subsection{OPE Side}

On OPE side,  the aforementioned correlation function is calculated in terms of QCD degrees of freedom and photon DAs. For this aim, we substitute
the explicit forms of the interpolating currents of the heavy
baryons  into the
correlation function in Eq. (\ref{T}) and use the Wick's theorem to obtain the correlation in terms of quark propagators.

The interpolating currents  for  spin 3/2  baryons are taken as
%
\begin{eqnarray}\label{currentguy}
\eta_{\mu}=A\epsilon^{abc}\left\{\vphantom{\int_0^{x_2}}(q_{1}^{a}C\gamma_{\mu}q_{2}^{b})Q^{c}
+(q_{2}^{a}C\gamma_{\mu}Q^{b})q_{1}^{c}+
(Q^{a}C\gamma_{\mu}q_{1}^{b})q_{2}^{c}\right\},
\end{eqnarray}
where $q_{1}$ and $q_{2}$ stand for light quarks; $a$, $b$ and $c$ are  color indices and $C$ is the charge
conjugation operator. The  normalization factor $A$ and light quark content of heavy spin 3/2 baryons are
presented in table \ref{tabel1}:
\begin{table}[h]
\centering
\begin{tabular}{|c||c|c|c|}\hline
  Heavy spin 3/2 baryons&$A$ & $q_{1}$& $q_{2}$\\\cline{1-4}
\hline\hline $\Omega_{b(c)}^{*-(0)}$
&$1/\sqrt{3}$&s&s\\\cline{1-4}
 $\Xi_{b(c)}^{*0(+)}$&$\sqrt{2/3}$&s&u\\\cline{1-4}
 $\Xi_{b(c)}^{*-(0)}$&$\sqrt{2/3}$&s&d\\\cline{1-4}
 \end{tabular}
 \vspace{0.8cm}
\caption{The  normalization factor $A$ and light quark content of heavy spin 3/2 baryons.
}\label{tabel1}
\end{table}

 The general
form of the interpolating currents for the heavy spin 1/2 baryons under consideration can be written as (see for instance \cite{Bagan})
\begin{eqnarray}
\label{estp05}
\eta \es - {B\over \sqrt{2}} \epsilon^{abc} \Big\{ \Big( q_1^{aT}
C Q^b \Big) \gamma_5 q_2^c + \beta \Big( q_1^{aT} C \gamma_5 Q^b \Big) q_2^c -
\Big[\Big( Q^{aT} C q_2^b \Big) \gamma_5 q_1^c + \beta \Big( Q^{aT} C
\gamma_5 q_2^b \Big) q_1^c \Big] \Big\}~, \nnb\\
\end{eqnarray}
where $\beta$ is an arbitrary parameter and $\beta=-1$ corresponds
to the Ioffe current. The constant $B$ and quark fields $q_{1}$ and $q_{2}$
 for the corresponding heavy spin 1/2 baryons are given in table \ref{tabel2}.
\begin{table}[h]
\centering
\begin{tabular}{|c||c|c|c|}\hline
  Heavy spin 3/2 baryons&$B$ & $q_{1}$& $q_{2}$\\\cline{1-4}
\hline\hline $\Omega_{b(c)}^{-(0)}$
&$1/\sqrt{2}$&s&s\\\cline{1-4}
 $\Xi_{b(c)}^{'0(+)}$&$1$&s&u\\\cline{1-4}
 $\Xi_{b(c)}^{'-(0)}$&$1$&s&d\\\cline{1-4}
 \end{tabular}
 \vspace{0.8cm}
\caption{The  constant $B$ and light quark content of the heavy spin 1/2 baryons under consideration.
}\label{tabel2}
\end{table}
%

The correlation function in OPE side  receives three
different contributions: 1) perturbative contributions, 2) mixed
contributions at which the photon is radiated from  short distances
and at least one of the quarks forms a condensate and  3)
non-perturbative contributions where photon is radiated at
long distances. The last contribution is parameterized by the matrix
element $ \langle\gamma(q)\mid\bar q(x_{1})
 \Gamma q(x_{2})\mid0\rangle$ which is expanded in terms of photon
 DAs with definite twists. Here $\Gamma$ is the full set of Dirac matrices
$\Gamma_j = \Big\{ 1,~\gamma_5,~\gamma_\alpha,~i\gamma_5
\gamma_\alpha, ~\sigma_{\alpha \beta} /\sqrt{2}\Big\}$.

The perturbative  contribution at which the photon interacts
with the quarks perturbatively, is obtained by  replacing corresponding free quark propagator by
\begin{equation}\label{rep1guy}
S^{ab}_{\alpha \beta} \Rightarrow  \left\{ \int d^4 y
S^{free} (x-y) \not\!A
S^{free}(y)\right\}^{ab}_{\alpha \beta},
\end{equation}
 where the free   light and heavy quark propagators are given as
\begin{eqnarray}\label{free1guy}
S^{free}_{q} &=&\frac{i\not\!x}{2\pi^{2}x^{4}}-\frac{m_{q}}{4\pi^{2}x^{2}},\nonumber\\
S^{free}_{Q}
&=&\frac{m_{Q}^{2}}{4\pi^{2}}\frac{K_{1}(m_{Q}\sqrt{-x^2})}{\sqrt{-x^2}}-i
\frac{m_{Q}^{2}\not\!x}{4\pi^{2}x^2}K_{2}(m_{Q}\sqrt{-x^2}),\nonumber\\
\end{eqnarray}
with $K_{i}$ being the  Bessel functions.

The non-perturbative contributions are obtained by  replacing one of the light quark
propagators that emits a photon by
\begin{equation}\label{rep2guy}
\label{rep} S^{ab}_{\alpha \beta} \rightarrow - \frac{1}{4} \bar q^a \Gamma_j q^b ( \Gamma_j )_{\alpha \beta}~,
\end{equation}
where sum
over  $j$ is applied, and the remaining by full quark propagators   involving the perturbative as well as the
non-perturbative parts. The full heavy and light quark propagators which we use in the present work are  (see \cite{Balitsky,Braun2})
\begin{eqnarray}\label{heavylightguy}
 S_Q (x)& =&  S_Q^{free} (x) - i g_s \int \frac{d^4 k}{(2\pi)^4}
e^{-ikx} \int_0^1 dv \Bigg[\frac{\not\!k + m_Q}{( m_Q^2-k^2)^2}
G^{\mu\nu}(vx)
\sigma_{\mu\nu} \nonumber \\
&+& \frac{1}{m_Q^2-k^2} v x_\mu G^{\mu\nu} \gamma_\nu \Bigg],
\nonumber \\
S_q(x) &=&  S_q^{free} (x) - \frac{m_q}{4 \pi^2 x^2} - \frac{\langle
\bar q q \rangle}{12} \left(1 - i \frac{m_q}{4} \not\!x \right) -
\frac{x^2}{192} m_0^2 \langle \bar q q \rangle \left( 1 - i
\frac{m_q}{6}\not\!x \right) \nonumber \\ &&
 - i g_s \int_0^1 du \left[\frac{\not\!x}{16 \pi^2 x^2} G_{\mu \nu} (ux) \sigma_{\mu \nu} - u x^\mu
G_{\mu \nu} (ux) \gamma^\nu \frac{i}{4 \pi^2 x^2} \right. \nonumber
\\ && \left. - i \frac{m_q}{32 \pi^2} G_{\mu \nu} \sigma^{\mu \nu}
\left( \ln \left( \frac{-x^2 \Lambda^2}{4} \right) + 2 \gamma_E
\right) \right],
 \end{eqnarray}
  where $\Lambda$ is the scale parameter and we choose it at factorization scale
  $\Lambda=(0.5-1)~GeV$  \cite{revised2}.

In order to calculate the non-perturbative contributions, we need
the matrix elements  $ \langle\gamma(q)\mid\bar q
 \Gamma_{i}q\mid0\rangle$. These matrix  elements are determined  in terms of
 the photon DAs as  \cite{Ball}
 \begin{eqnarray}
&&\langle \gamma(q) \vert  \bar q(x) \sigma_{\mu \nu} q(0) \vert  0
\rangle  = -i e_q \bar q q (\varepsilon_\mu q_\nu - \varepsilon_\nu
q_\mu) \int_0^1 du e^{i \bar u qx} \left(\chi \varphi_\gamma(u) +
\frac{x^2}{16} \mathbb{A}  (u) \right) \nonumber \\ &&
-\frac{i}{2(qx)}  e_q \qq \left[x_\nu \left(\varepsilon_\mu - q_\mu
\frac{\varepsilon x}{qx}\right) - x_\mu \left(\varepsilon_\nu -
q_\nu \frac{\varepsilon x}{q x}\right) \right] \int_0^1 du e^{i \bar
u q x} h_\gamma(u),
\nonumber \\
&&\langle \gamma(q) \vert  \bar q(x) \gamma_\mu q(0) \vert 0 \rangle
= e_q f_{3 \gamma} \left(\varepsilon_\mu - q_\mu \frac{\varepsilon
x}{q x} \right) \int_0^1 du e^{i \bar u q x} \psi^v(u),
\nonumber \\
&&\langle \gamma(q) \vert \bar q(x) \gamma_\mu \gamma_5 q(0) \vert 0
\rangle  = - \frac{1}{4} e_q f_{3 \gamma} \epsilon_{\mu \nu \alpha
\beta } \varepsilon^\nu q^\alpha x^\beta \int_0^1 du e^{i \bar u q
x} \psi^a(u),
\nonumber \\
&&\langle \gamma(q) | \bar q(x) g_s G_{\mu \nu} (v x) q(0) \vert 0
\rangle = -i e_q \qq \left(\varepsilon_\mu q_\nu - \varepsilon_\nu
q_\mu \right) \int {\cal D}\alpha_i e^{i (\alpha_{\bar q} + v
\alpha_g) q x} {\cal S}(\alpha_i),
\nonumber \\
&&\langle \gamma(q) | \bar q(x) g_s \tilde G_{\mu \nu} i \gamma_5 (v
x) q(0) \vert 0 \rangle = -i e_q \qq \left(\varepsilon_\mu q_\nu -
\varepsilon_\nu q_\mu \right) \int {\cal D}\alpha_i e^{i
(\alpha_{\bar q} + v \alpha_g) q x} \tilde {\cal S}(\alpha_i),
\nonumber \\
&&\langle \gamma(q) \vert \bar q(x) g_s \tilde G_{\mu \nu}(v x)
\gamma_\alpha \gamma_5 q(0) \vert 0 \rangle = e_q f_{3 \gamma}
q_\alpha (\varepsilon_\mu q_\nu - \varepsilon_\nu q_\mu) \int {\cal
D}\alpha_i e^{i (\alpha_{\bar q} + v \alpha_g) q x} {\cal
A}(\alpha_i),
\nonumber \\
&&\langle \gamma(q) \vert \bar q(x) g_s G_{\mu \nu}(v x) i
\gamma_\alpha q(0) \vert 0 \rangle = e_q f_{3 \gamma} q_\alpha
(\varepsilon_\mu q_\nu - \varepsilon_\nu q_\mu) \int {\cal
D}\alpha_i e^{i (\alpha_{\bar q} + v \alpha_g) q x} {\cal
V}(\alpha_i) ,\nonumber \\ && \langle \gamma(q) \vert \bar q(x)
\sigma_{\alpha \beta} g_s G_{\mu \nu}(v x) q(0) \vert 0 \rangle  =
e_q \qq \left\{
        \left[\left(\varepsilon_\mu - q_\mu \frac{\varepsilon x}{q x}\right)\left(g_{\alpha \nu} -
        \frac{1}{qx} (q_\alpha x_\nu + q_\nu x_\alpha)\right) \right. \right. q_\beta
\nonumber \\ && -
         \left(\varepsilon_\mu - q_\mu \frac{\varepsilon x}{q x}\right)\left(g_{\beta \nu} -
        \frac{1}{qx} (q_\beta x_\nu + q_\nu x_\beta)\right) q_\alpha
\nonumber \\ && -
         \left(\varepsilon_\nu - q_\nu \frac{\varepsilon x}{q x}\right)\left(g_{\alpha \mu} -
        \frac{1}{qx} (q_\alpha x_\mu + q_\mu x_\alpha)\right) q_\beta
\nonumber \\ &&+
         \left. \left(\varepsilon_\nu - q_\nu \frac{\varepsilon x}{q.x}\right)\left( g_{\beta \mu} -
        \frac{1}{qx} (q_\beta x_\mu + q_\mu x_\beta)\right) q_\alpha \right]
   \int {\cal D}\alpha_i e^{i (\alpha_{\bar q} + v \alpha_g) qx} {\cal T}_1(\alpha_i)
\nonumber \\ &&+
        \left[\left(\varepsilon_\alpha - q_\alpha \frac{\varepsilon x}{qx}\right)
        \left(g_{\mu \beta} - \frac{1}{qx}(q_\mu x_\beta + q_\beta x_\mu)\right) \right. q_\nu
\nonumber \\ &&-
         \left(\varepsilon_\alpha - q_\alpha \frac{\varepsilon x}{qx}\right)
        \left(g_{\nu \beta} - \frac{1}{qx}(q_\nu x_\beta + q_\beta x_\nu)\right)  q_\mu
\nonumber \\ && -
         \left(\varepsilon_\beta - q_\beta \frac{\varepsilon x}{qx}\right)
        \left(g_{\mu \alpha} - \frac{1}{qx}(q_\mu x_\alpha + q_\alpha x_\mu)\right) q_\nu
\nonumber \\ &&+
         \left. \left(\varepsilon_\beta - q_\beta \frac{\varepsilon x}{qx}\right)
        \left(g_{\nu \alpha} - \frac{1}{qx}(q_\nu x_\alpha + q_\alpha x_\nu) \right) q_\mu
        \right]
    \int {\cal D} \alpha_i e^{i (\alpha_{\bar q} + v \alpha_g) qx} {\cal T}_2(\alpha_i)
\nonumber \\ &&+
        \frac{1}{qx} (q_\mu x_\nu - q_\nu x_\mu)
        (\varepsilon_\alpha q_\beta - \varepsilon_\beta q_\alpha)
    \int {\cal D} \alpha_i e^{i (\alpha_{\bar q} + v \alpha_g) qx} {\cal T}_3(\alpha_i)
\nonumber \\ &&+
        \left. \frac{1}{qx} (q_\alpha x_\beta - q_\beta x_\alpha)
        (\varepsilon_\mu q_\nu - \varepsilon_\nu q_\mu)
    \int {\cal D} \alpha_i e^{i (\alpha_{\bar q} + v \alpha_g) qx} {\cal T}_4(\alpha_i)
                        \right\},
\end{eqnarray}
where
$\varphi_\gamma(u)$ is the leading twist 2, $\psi^v(u)$,
$\psi^a(u)$, ${\cal A}$ and ${\cal V}$ are the twist 3; and
$h_\gamma(u)$, $\mathbb{A}$ and ${\cal T}_i$ ($i=1,~2,~3,~4$) are the
twist 4 photon DAs \cite{Ball}. Here $\chi$ is the magnetic susceptibility of the quarks.

 The measure $\int{\cal D} \alpha_i$ is defined as
\begin{equation}
\int {\cal D} \alpha_i = \int_0^1 d \alpha_{\bar q} \int_0^1 d
\alpha_q \int_0^1 d \alpha_g \delta(1-\alpha_{\bar
q}-\alpha_q-\alpha_g).\nonumber \\
\end{equation}
In order to obtain the sum rules for the form factors $G_{1}$ and
$G_{2}$, we equate the coefficients of the structures
$\not\!\varepsilon \not\!p\gamma_{5}q_{\mu}$ and
$\not\!q\not\!p\gamma_{5}(\varepsilon p)q_{\mu}$  from
both hadronic and OPE representations of the same correlation function. We apply the Borel
transformations with respect to the variables $p^2$ and $(p+q)^2$
as well as continuum subtraction to suppress the contributions of the higher states and continuum.
Finally, we obtain the following schematically written  sum rules for the
electromagnetic form factors $G_{1}$  and $G_{2}$ :
\begin{eqnarray}\label{magneticmoment1}
G_{1}&=&-\frac{1}{\lambda_{1}\lambda_{2}(m_{1}+m_{2})}e^{\frac{m_{1}^{2}}{M_{1}^{2}}}
e^{\frac{m_{2}^{2}}{M_{2}^{2}}}
\left[\vphantom{\int_0^{x_2}}e_{q_{1}}\Pi_{1}+e_{q_{2}}\Pi_{1}(q_{1}\leftrightarrow
q_{2})+e_{Q}\Pi'_{1}\right]
\nonumber\\
G_{2}&=&\frac{1}{\lambda_{1}\lambda_{2}}e^{\frac{m_{1}^{2}}{M_{1}^{2}}}e^{\frac{m_{2}^{2}}
{M_{2}^{2}}}
\left[\vphantom{\int_0^{x_2}}e_{q_{1}}\Pi_{2}+e_{q_{2}}\Pi_{2}(q_{1}\leftrightarrow q_{2})
+e_{Q}\Pi'_{2}\right],
\end{eqnarray}
where the functions $\Pi_i[\Pi'_{i}]$ can be written as
\begin{eqnarray}\label{magneticmoment2}
\Pi_{i}[\Pi'_{i}]&=&\int_{m_{Q}^{2}}^{s_{0}}e^{\frac{-s}{M^{2}}}\rho_{i}(s)[\rho'_{i}(s)]ds
+e^{\frac{-m_Q^2}{M^{2}}}
\Gamma_{i}[\Gamma'_{i}],
\end{eqnarray}
where $s_0$ is the continuum threshold and we take $M_1^2=M_2^2= 2M^2$ since the masses of the initial and final baryons are  close to each other. The spectral densities $\rho_{i}(s)[\rho'_{i}(s)]$
and the functions $\Gamma_{i}[\Gamma'_{i}]$ are very lengthy, hence, we do not present their explicit expressions here.

\section{Numerical results}

In this part, we numerically analyze the sum rules for
the magnetic dipole $G_{M}$ and electric quadrupole  $G_{E}$ obtained in the previous section. For this aim, we use the input parameters
 $\uu(1~GeV) = \dd(1~GeV)= -(0.243)^3~GeV^3$, $\ss(1~GeV) = 0.8
\uu(1~GeV)$, $m_0^2(1~GeV) = (0.8\pm0.2)~GeV^2$ \cite{Belyaev} and $f_{3 \gamma} = - 0.0039~GeV^2$ \cite{Ball}. The value
of the magnetic susceptibility  are
calculated   in    \cite{Rohrwild,balitskibal,Kogan}. Here we use the value   $\chi(1~GeV)=-4.4~GeV^{-2}$ \cite{Kogan} for this quantity.  The LCQSR for the
 magnetic dipole
and electric quadrupole  moments also include the
  photon DAs \cite{Ball} whose expressions are given as
\begin{eqnarray}
\varphi_\gamma(u) &=& 6 u \bar u \left( 1 + \varphi_2(\mu)
C_2^{\frac{3}{2}}(u - \bar u) \right),
\nonumber \\
\psi^v(u) &=& 3 \left(3 (2 u - 1)^2 -1 \right)+\frac{3}{64} \left(15
w^V_\gamma - 5 w^A_\gamma\right)
                        \left(3 - 30 (2 u - 1)^2 + 35 (2 u -1)^4
                        \right),
\nonumber \\
\psi^a(u) &=& \left(1- (2 u -1)^2\right)\left(5 (2 u -1)^2 -1\right)
\frac{5}{2}
    \left(1 + \frac{9}{16} w^V_\gamma - \frac{3}{16} w^A_\gamma
    \right),
\nonumber \\
{\cal A}(\alpha_i) &=& 360 \alpha_q \alpha_{\bar q} \alpha_g^2
        \left(1 + w^A_\gamma \frac{1}{2} (7 \alpha_g - 3)\right),
\nonumber \\
{\cal V}(\alpha_i) &=& 540 w^V_\gamma (\alpha_q - \alpha_{\bar q})
\alpha_q \alpha_{\bar q}
                \alpha_g^2,
\nonumber \\
h_\gamma(u) &=& - 10 \left(1 + 2 \kappa^+\right) C_2^{\frac{1}{2}}(u
- \bar u),
\nonumber \\
\mathbb{A}(u) &=& 40 u^2 \bar u^2 \left(3 \kappa - \kappa^+
+1\right) \nonumber \\ && +
        8 (\zeta_2^+ - 3 \zeta_2) \left[u \bar u (2 + 13 u \bar u) \right.
\nonumber \\ && + \left.
                2 u^3 (10 -15 u + 6 u^2) \ln(u) + 2 \bar u^3 (10 - 15 \bar u + 6 \bar u^2)
        \ln(\bar u) \right],
\nonumber \\
{\cal T}_1(\alpha_i) &=& -120 (3 \zeta_2 + \zeta_2^+)(\alpha_{\bar
q} - \alpha_q)
        \alpha_{\bar q} \alpha_q \alpha_g,
\nonumber \\
{\cal T}_2(\alpha_i) &=& 30 \alpha_g^2 (\alpha_{\bar q} - \alpha_q)
    \left((\kappa - \kappa^+) + (\zeta_1 - \zeta_1^+)(1 - 2\alpha_g) +
    \zeta_2 (3 - 4 \alpha_g)\right),
\nonumber \\
{\cal T}_3(\alpha_i) &=& - 120 (3 \zeta_2 - \zeta_2^+)(\alpha_{\bar
q} -\alpha_q)
        \alpha_{\bar q} \alpha_q \alpha_g,
\nonumber \\
{\cal T}_4(\alpha_i) &=& 30 \alpha_g^2 (\alpha_{\bar q} - \alpha_q)
    \left((\kappa + \kappa^+) + (\zeta_1 + \zeta_1^+)(1 - 2\alpha_g) +
    \zeta_2 (3 - 4 \alpha_g)\right),\nonumber \\
{\cal S}(\alpha_i) &=& 30\alpha_g^2\{(\kappa +
\kappa^+)(1-\alpha_g)+(\zeta_1 + \zeta_1^+)(1 - \alpha_g)(1 -
2\alpha_g)\nonumber \\&+&\zeta_2
[3 (\alpha_{\bar q} - \alpha_q)^2-\alpha_g(1 - \alpha_g)]\},\nonumber \\
\tilde {\cal S}(\alpha_i) &=&-30\alpha_g^2\{(\kappa -
\kappa^+)(1-\alpha_g)+(\zeta_1 - \zeta_1^+)(1 - \alpha_g)(1 -
2\alpha_g)\nonumber \\&+&\zeta_2 [3 (\alpha_{\bar q} -
\alpha_q)^2-\alpha_g(1 - \alpha_g)]\},
\end{eqnarray}
where, the constants inside the DAs are given by
 $\varphi_2(1~GeV) = 0$, $w^V_\gamma = 3.8 \pm 1.8$,
$w^A_\gamma = -2.1 \pm 1.0$, $\kappa = 0.2$, $\kappa^+ = 0$,
$\zeta_1 = 0.4$, $\zeta_2 = 0.3$, $\zeta_1^+ = 0$ and $\zeta_2^+ =
0$ \cite{Ball}.

The sum rules   for the electromagnetic form factors contain  three more auxiliary parameters:  Borel mass parameter $M^2$,
the continuum threshold $s_{0}$ and the arbitrary parameter
$\beta$  entering  the expressions of the interpolating currents of the
heavy spin 1/2 baryons. Any physical
quantities, like the magnetic dipole and  electric quadrupole
moments, should be independent of these auxiliary parameters.
Therefore, we try to find ``working regions"  for these auxiliary
parameters such that in these regions the $G_{M}$ and $G_{E}$ are
practically independent of these parameters. The upper and lower bands for
$M^2$ are found requiring that not only  the contributions of the
higher states and continuum are less than the ground state
contribution, but also the contributions of the higher twists are less compared to the leading twists. By these requirements, the working regions of Borel mass parameter are obtained as
$15~GeV^2\leq M^{2}\leq30~GeV^2 $ and $6~GeV^2\leq
M^{2}\leq12~GeV^2 $ for baryons containing b and c quarks,
respectively. The continuum threshold $s_{0}$ is the energy square which characterizes the beginning of the continuum. In we denote the ground state mass by $m$, the quantity $\sqrt{s_0}-m$
is the energy needed  to excite the particle to its first excited state with the same quantum numbers. The $\sqrt{s_0}-m$ is not well known for the baryons under consideration,
 but it should lie between $0.3~GeV$ and $0.8~GeV$.
The dependence of the magnetic dipole moment $G_{M}$ and electric quadrupole  moment $G_{E}$ on the Borel mass parameter
at different fixed values of the continuum threshold
and general parameter $\beta$ are depicted in figures 1-6 for the radiative transitions under consideration.
\begin{figure}[h!]
\includegraphics[width=8cm]{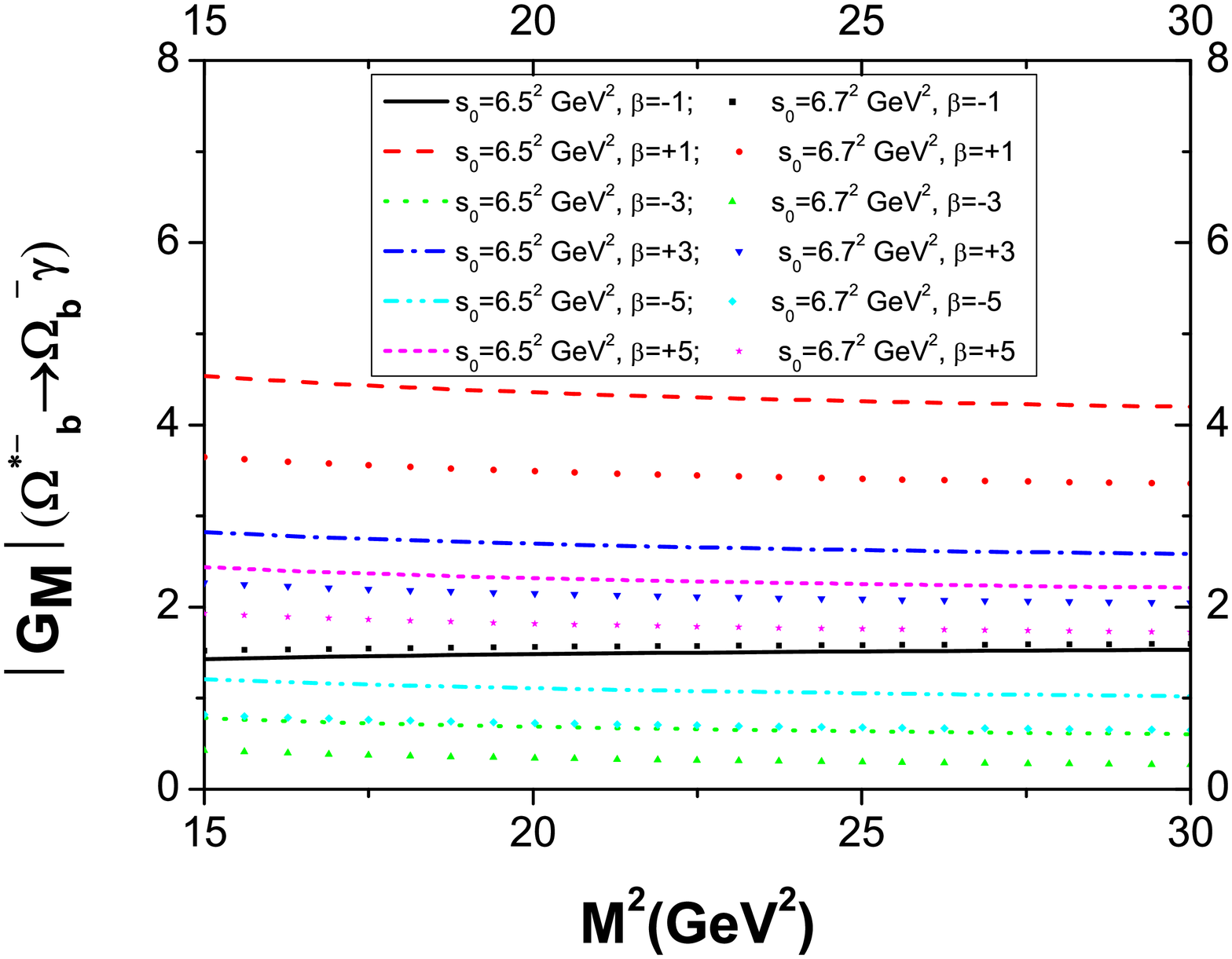}
\includegraphics[width=8cm]{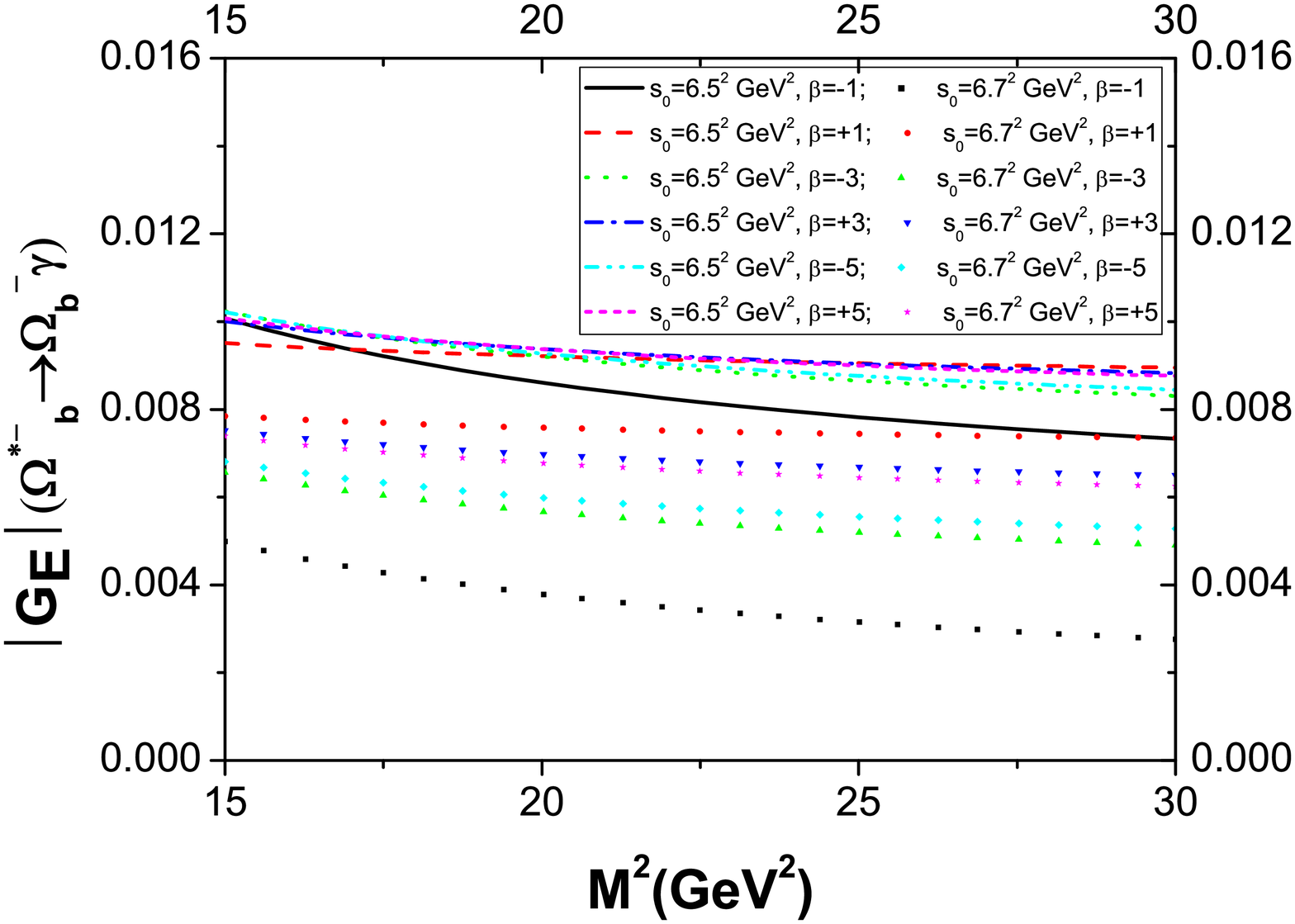}
\caption{ \textbf{Left:} The dependence of the magnetic dipole
moment $G_{M}$ for $\Omega^{*-}_{b}\rightarrow
\Omega^{-}_{b}\gamma$ transition on the Borel mass parameter $M^{2}$.
\textbf{Right:} The dependence of the electric quadrupole  moment
$G_{E}$ for $\Omega^{*-}_{b}\rightarrow \Omega^{-}_{b}\gamma$ transition on
the Borel mass parameter $M^{2}$. } \label{fig7a}
\end{figure}
\begin{figure}[h!]
\includegraphics[width=8cm]{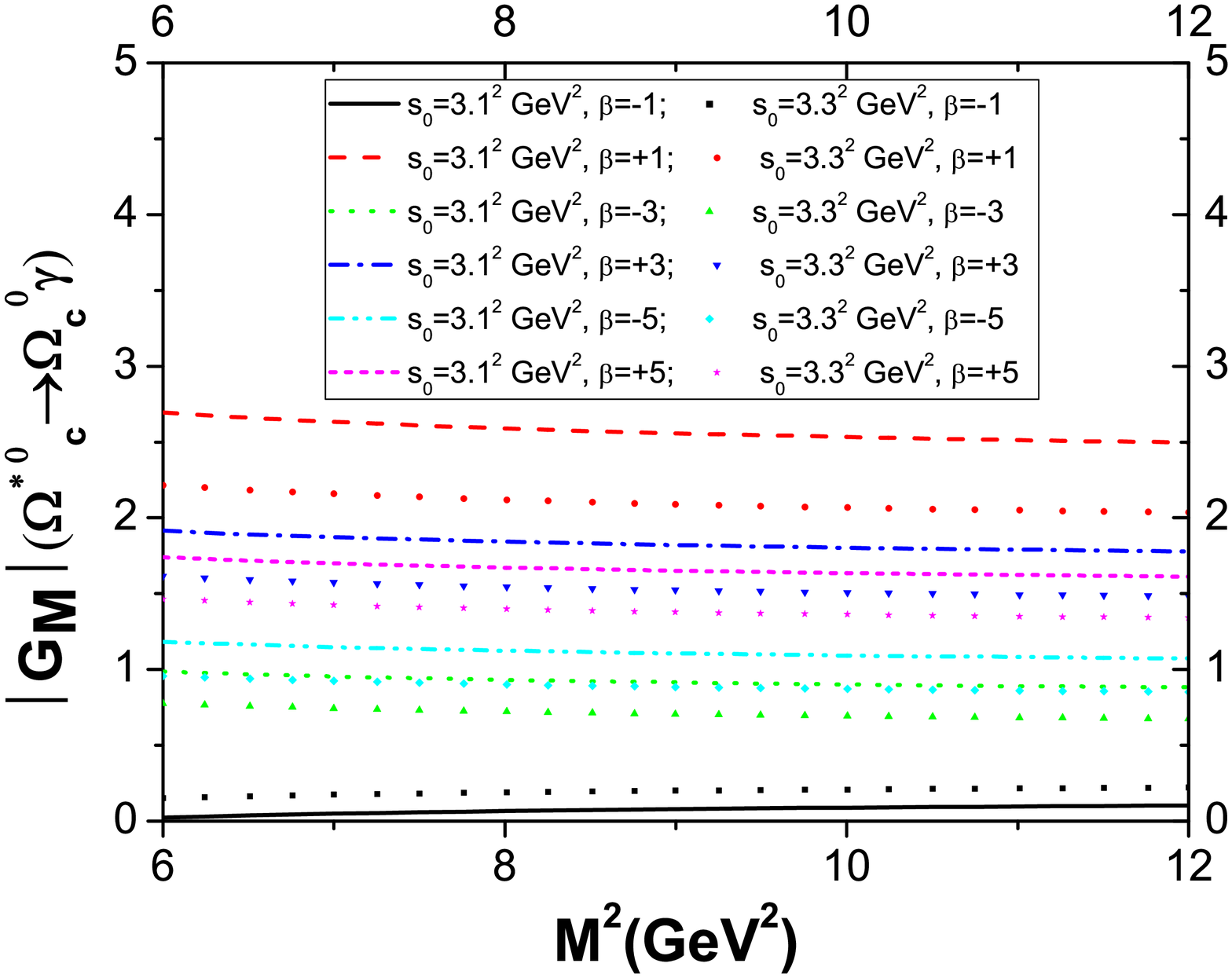}
\includegraphics[width=8cm]{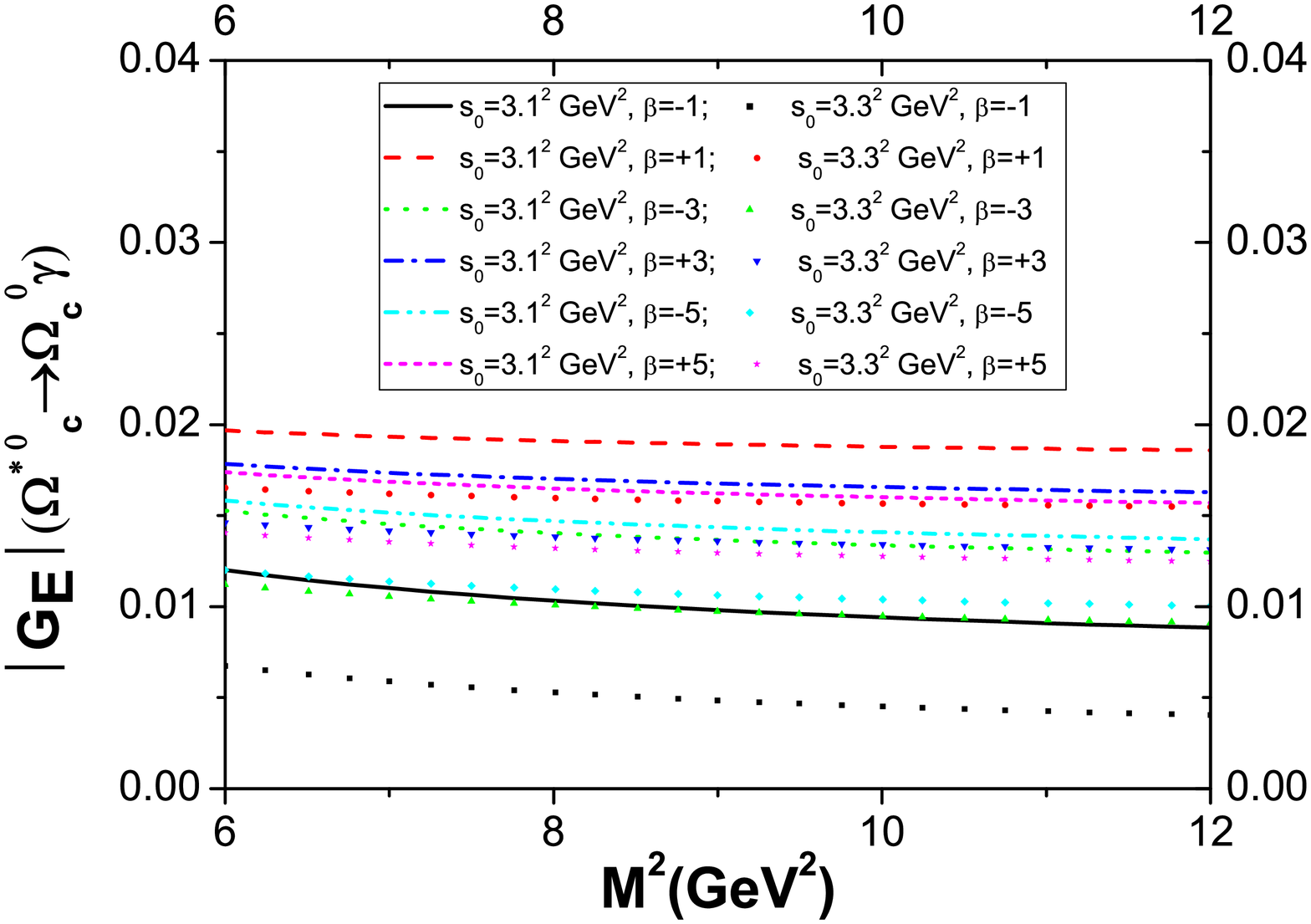}
\caption{The same as Fig. \ref{fig7a}, but  for
$\Omega^{*0}_{c}\rightarrow \Omega^{0}_{c}\gamma$.} \label{fig8a}
\end{figure}
\begin{figure}[h!]
\includegraphics[width=8cm]{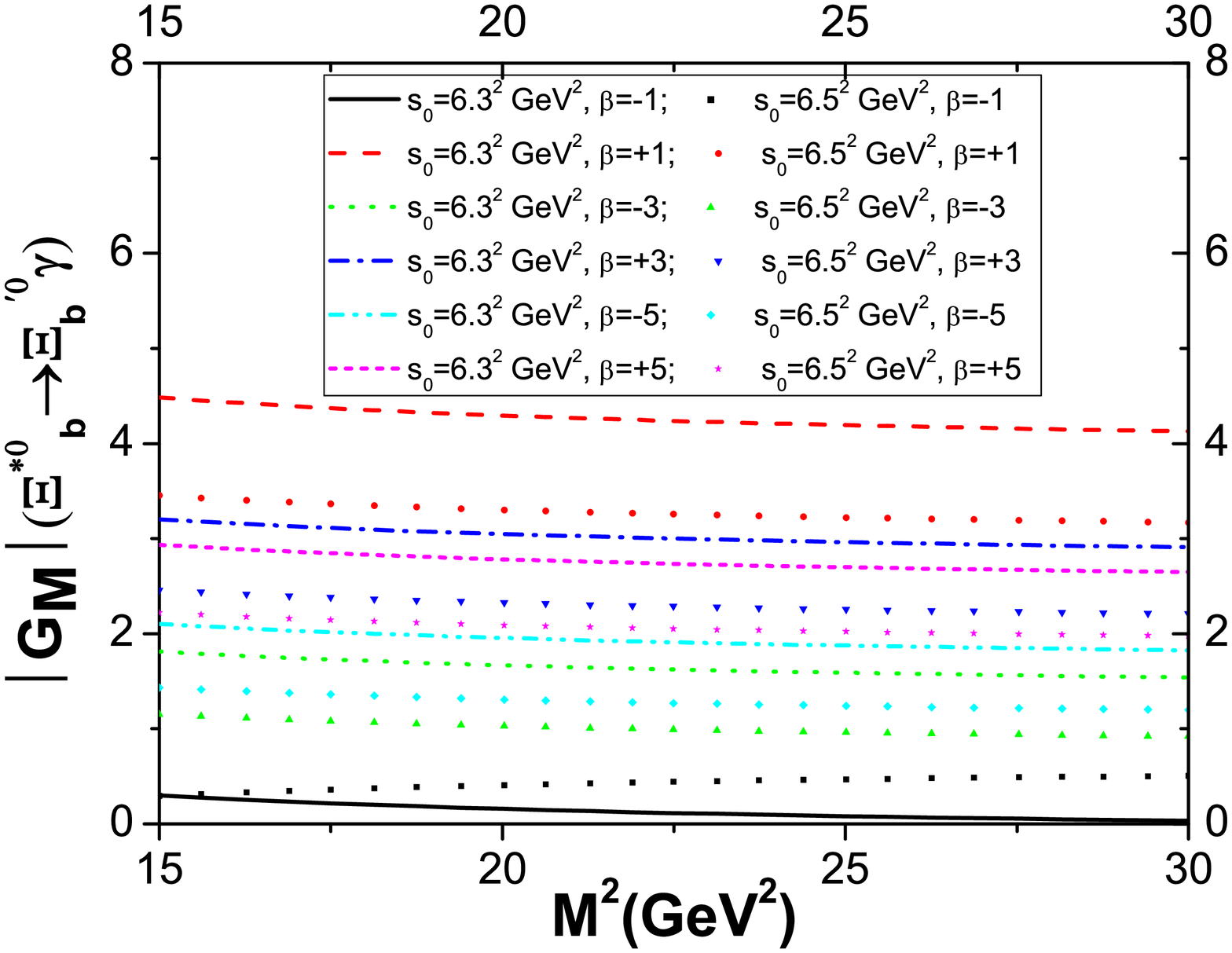}
\includegraphics[width=8cm]{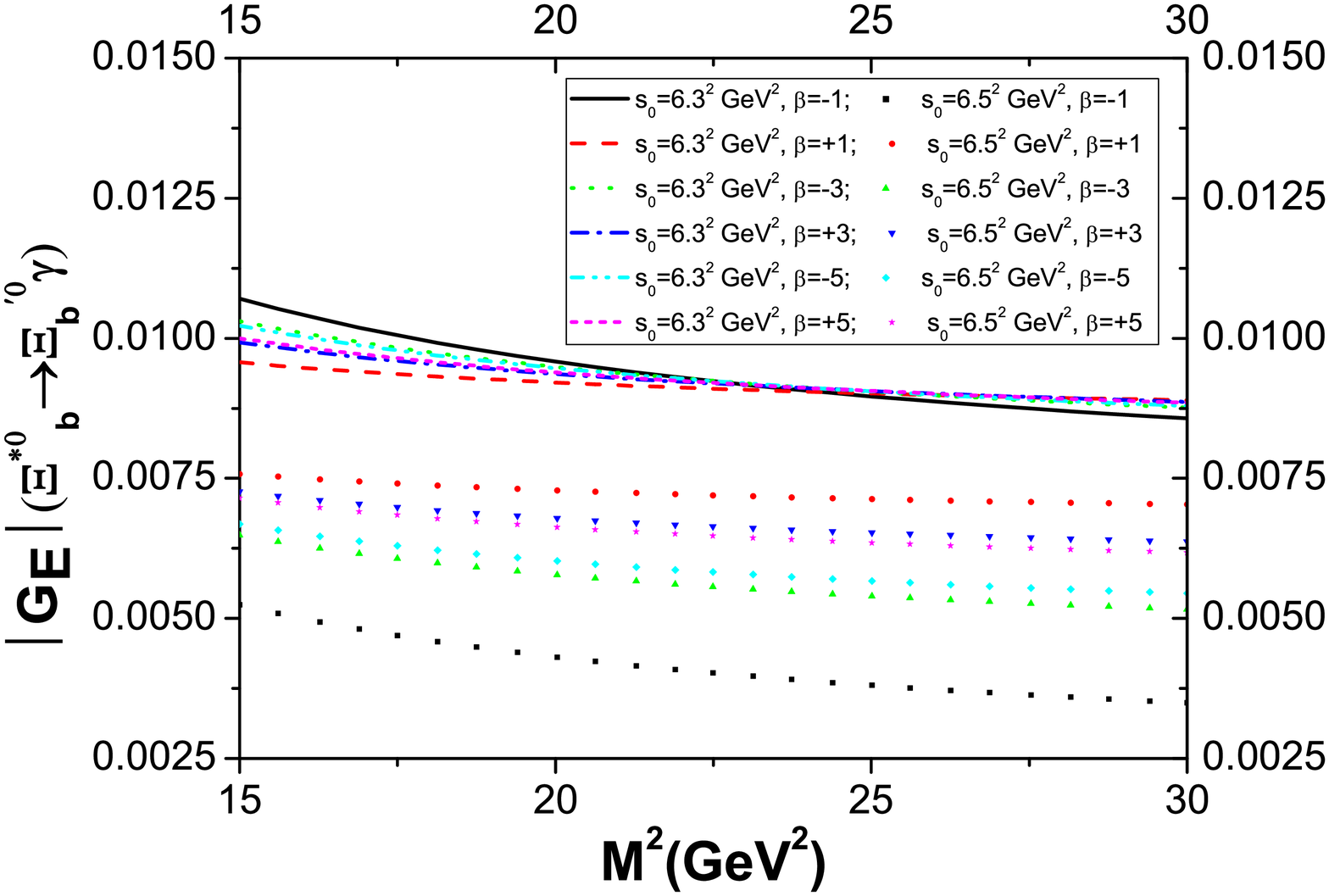}
\caption{The same as Fig. \ref{fig7a}, but  for
$\Xi^{*0}_{b}\rightarrow \Xi^{\prime 0}_{b}\gamma$.} \label{fig9a}
\end{figure}
\begin{figure}[h!]
\includegraphics[width=8cm]{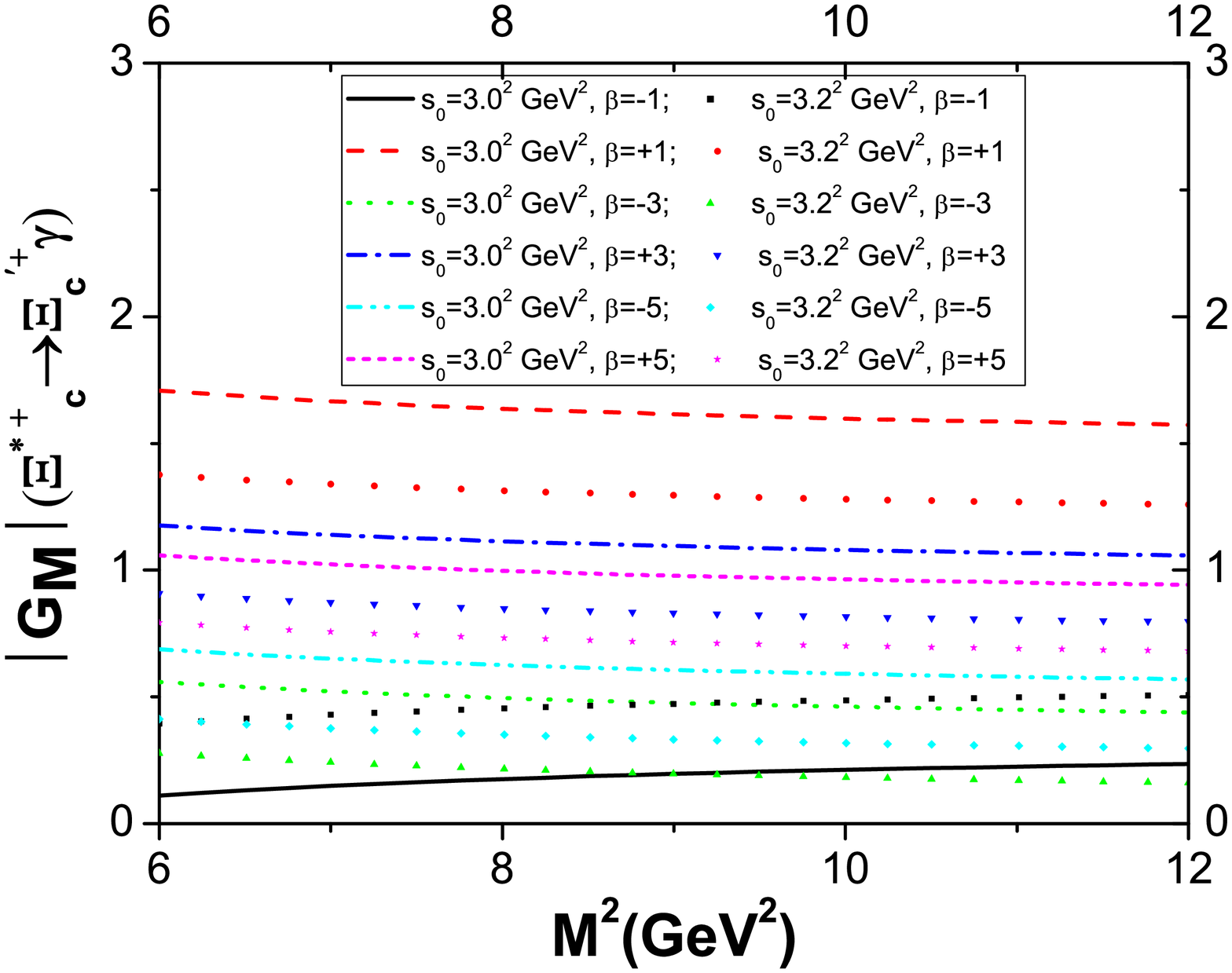}
\includegraphics[width=8cm]{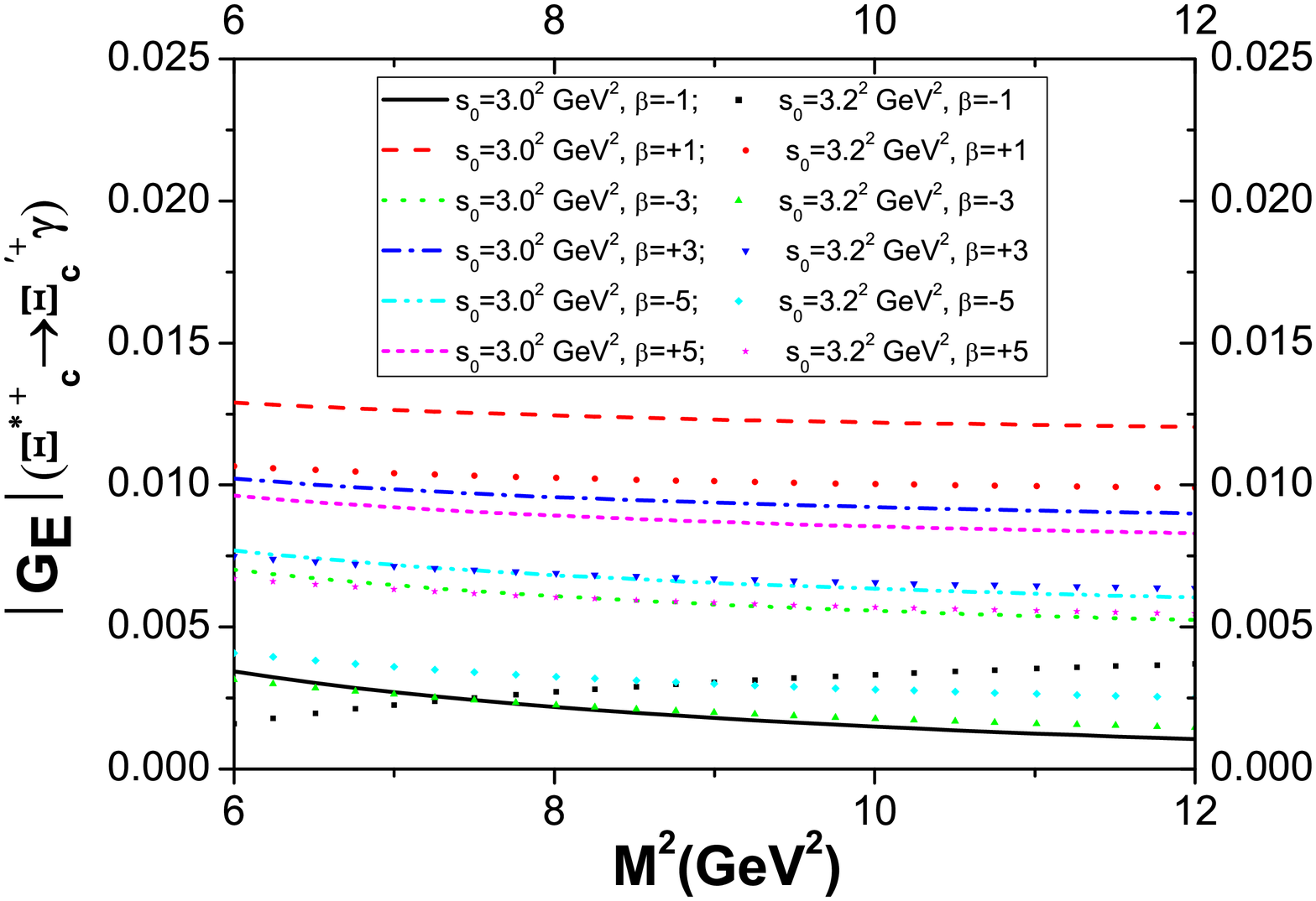}
\caption{The same as Fig. \ref{fig7a}, but  for
$\Xi^{*+}_{c}\rightarrow \Xi^{\prime +}_{c} \gamma$.}
\label{fig10a}
\end{figure}
\begin{figure}[h!]
\includegraphics[width=8cm]{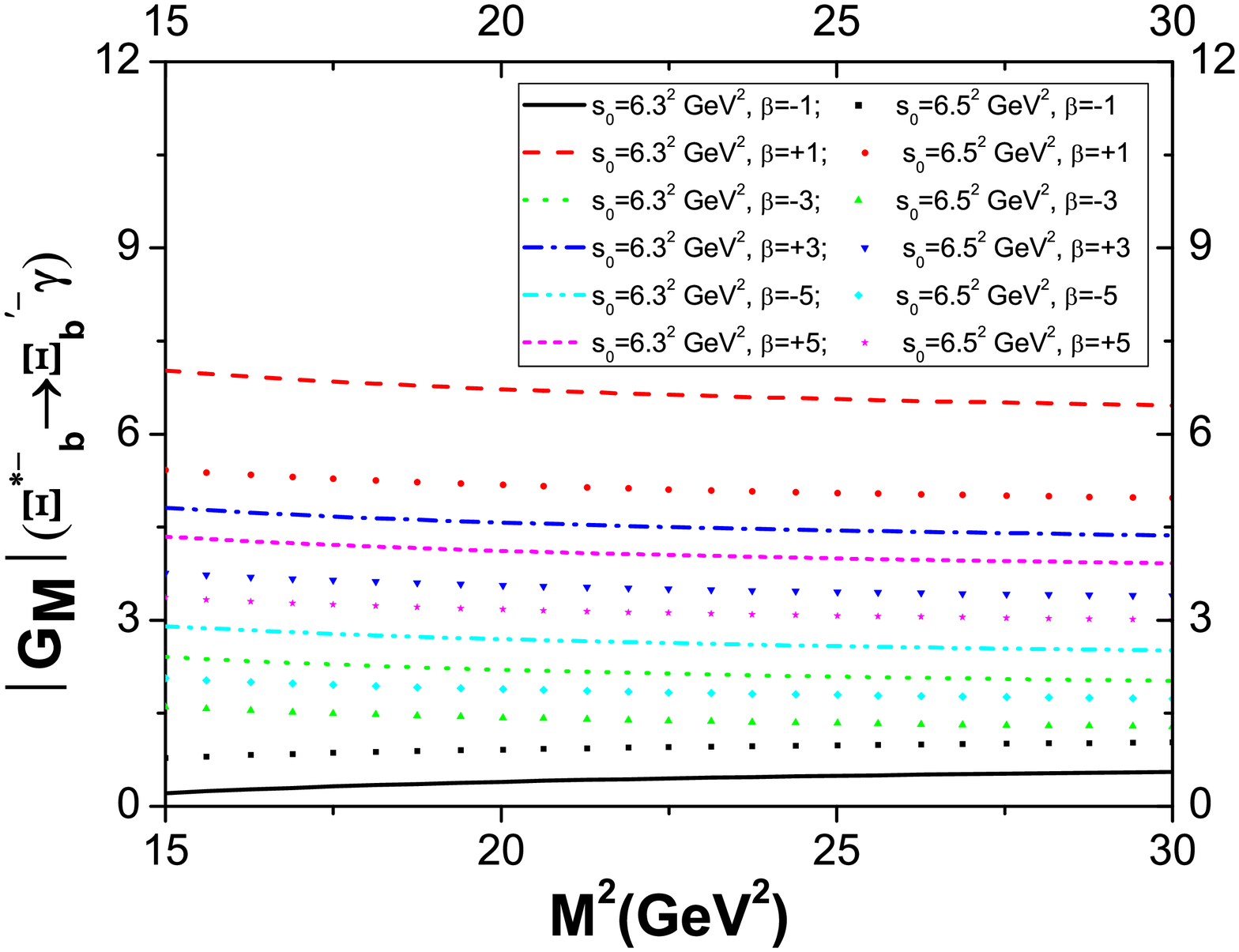}
\includegraphics[width=8cm]{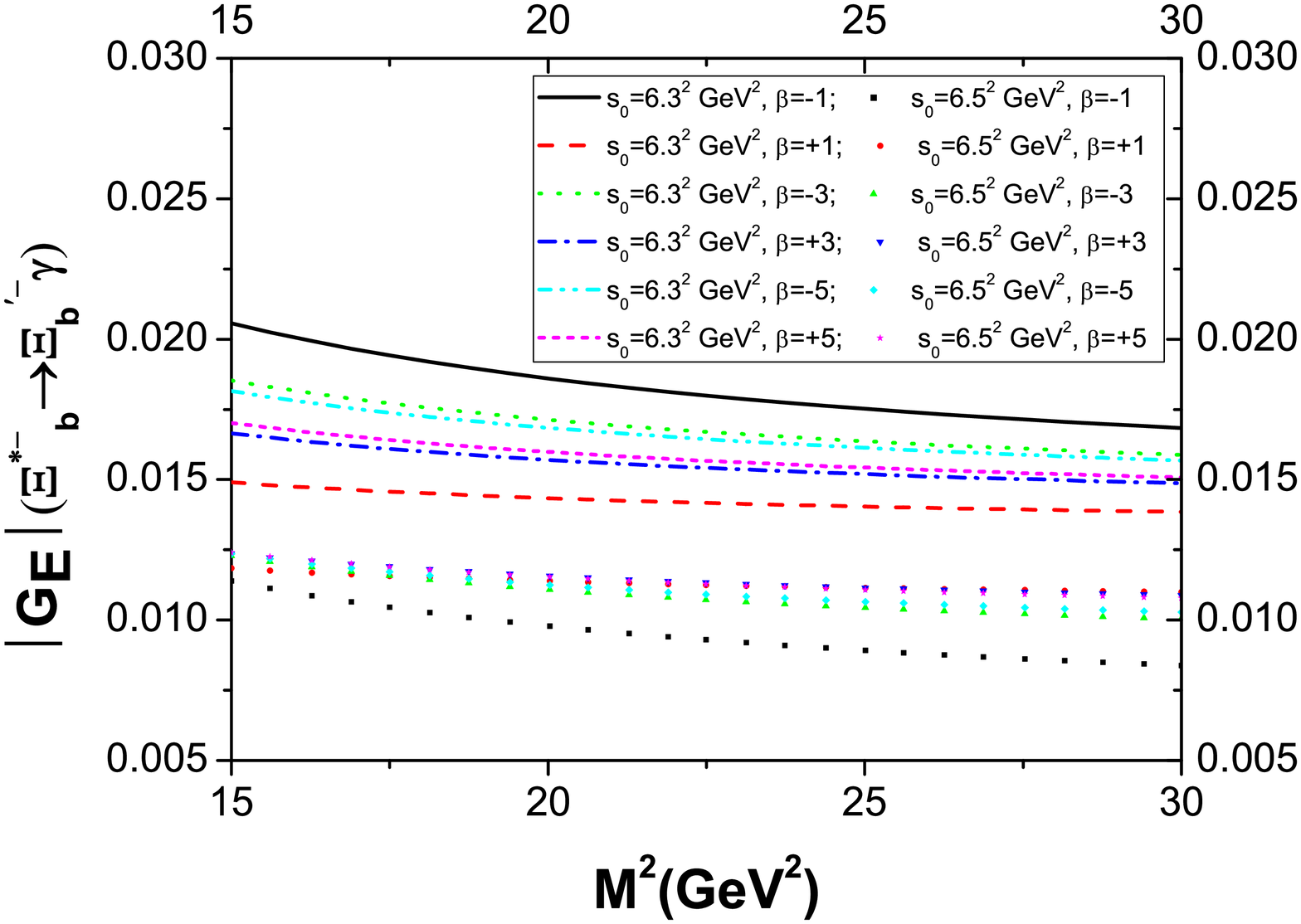}
\caption{The same as Fig. \ref{fig7a}, but  for
$\Xi^{*-}_{b}\rightarrow \Xi^{\prime -}_{b} \gamma$.}
\label{fig11a}
\end{figure}
\begin{figure}[h!]
\includegraphics[width=8cm]{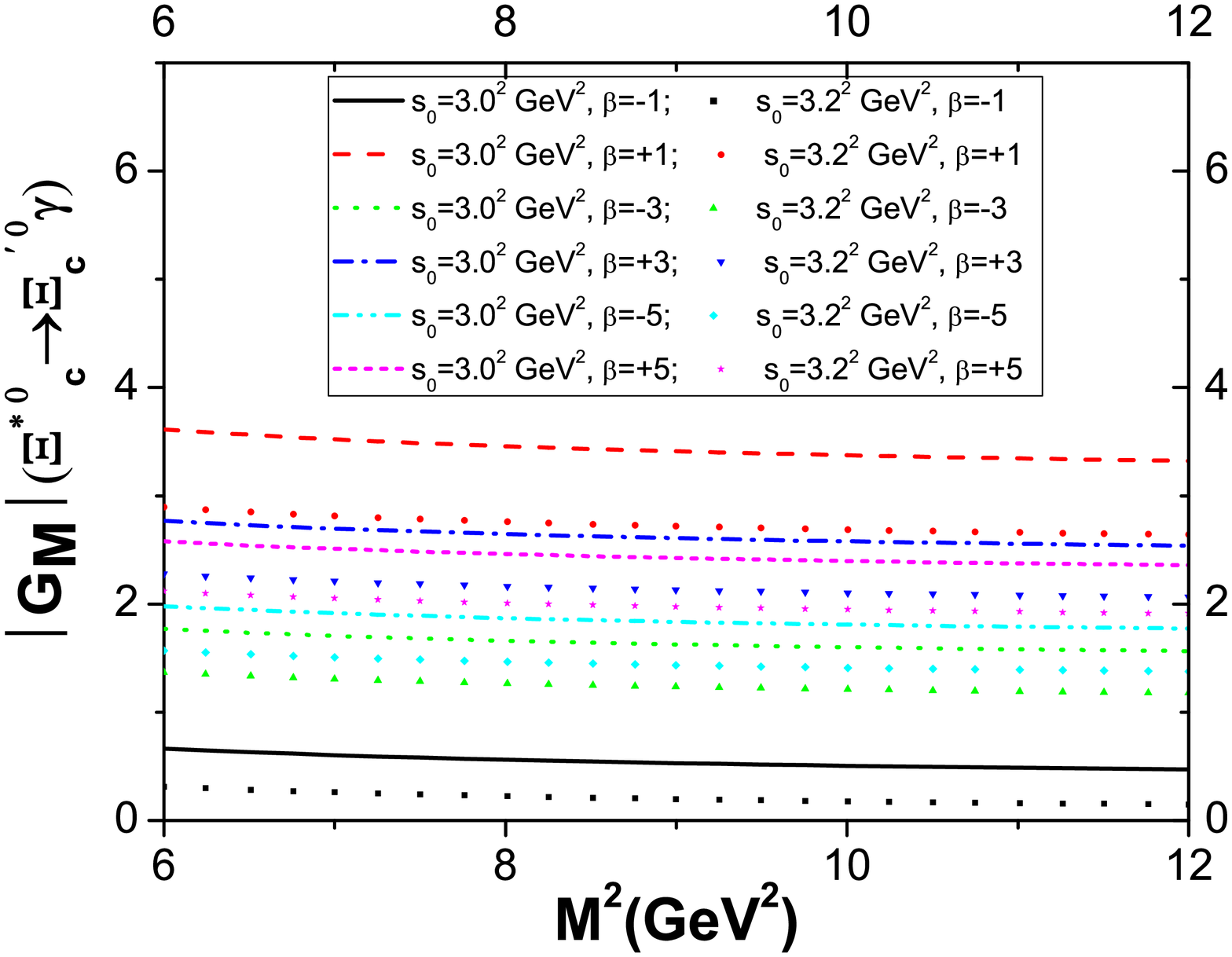}
\includegraphics[width=8cm]{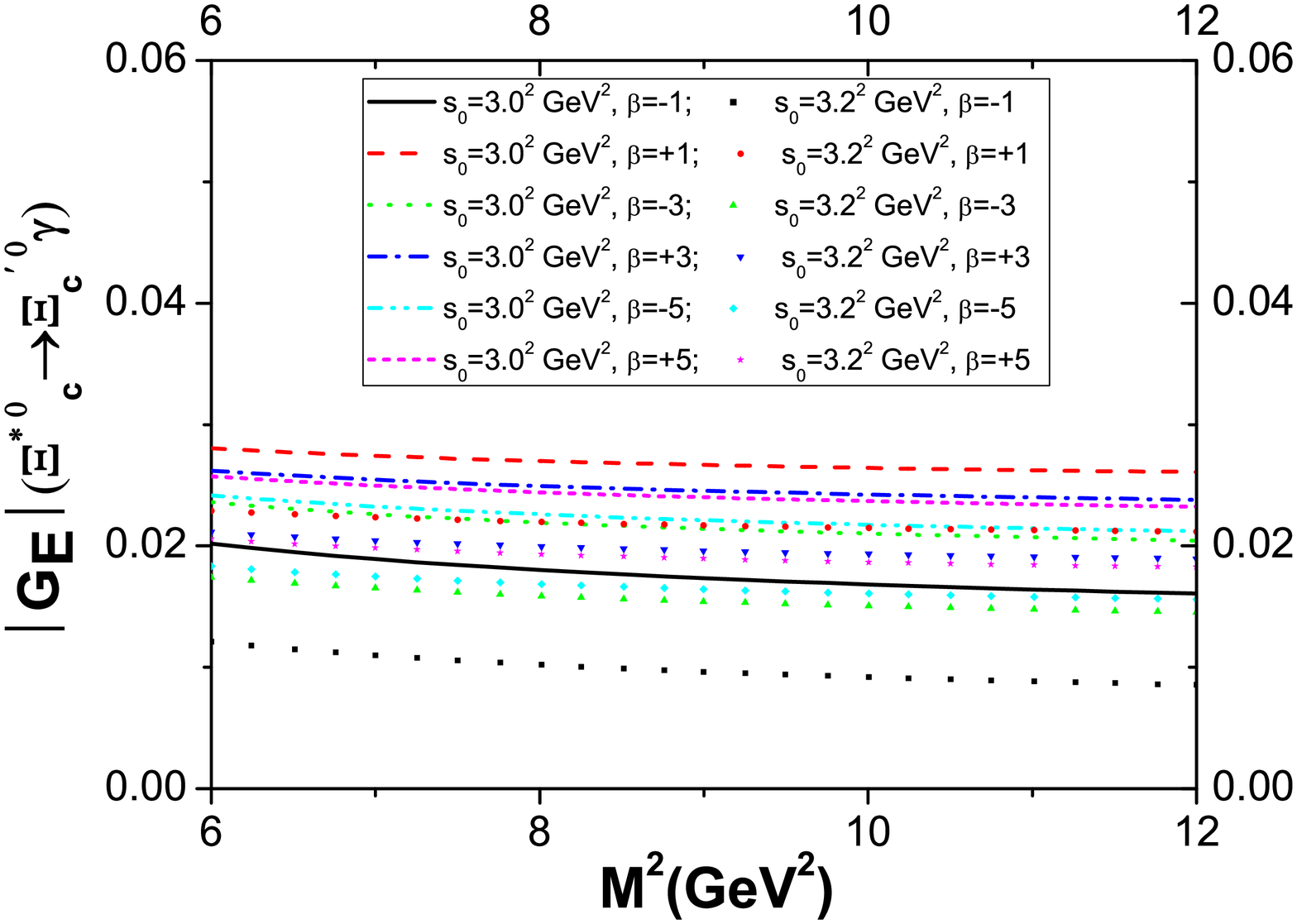}
\caption{The same as Fig. \ref{fig7a}, but  for
$\Xi^{*0}_{c}\rightarrow \Xi^{\prime 0}_{c} \gamma$.}
\label{fig12a}
\end{figure}
Note that, in all figures,  we plot  the absolute values of the physical quantities under study  since it
is not possible to predict the signs of the residues  from the mass
sum rules.   From these figures, we see that the results weakly depend on the $M^2$ and $s_0$ in their working regions.

\begin{figure}[h!]
\includegraphics[width=8cm]{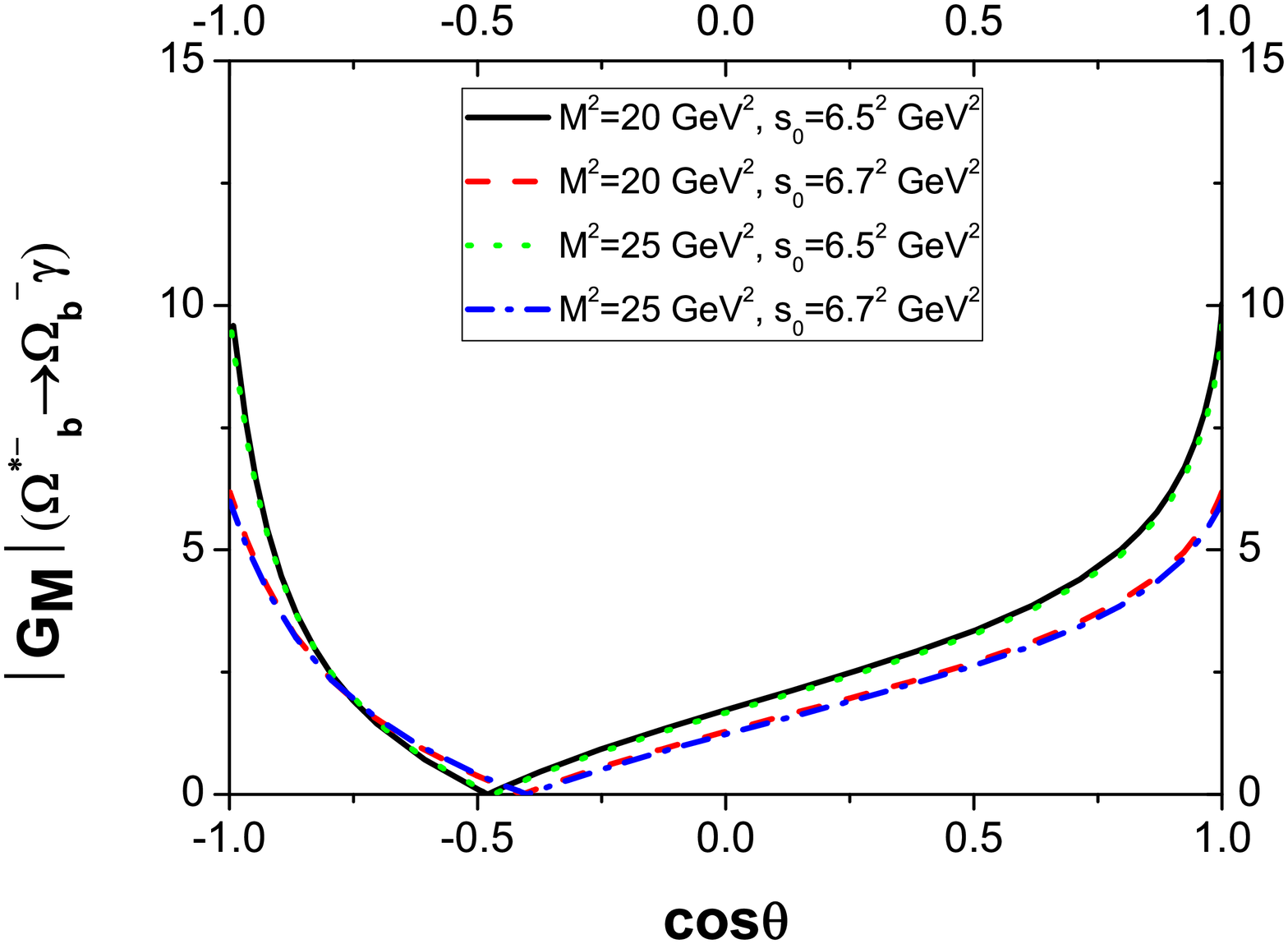}
\includegraphics[width=8cm]{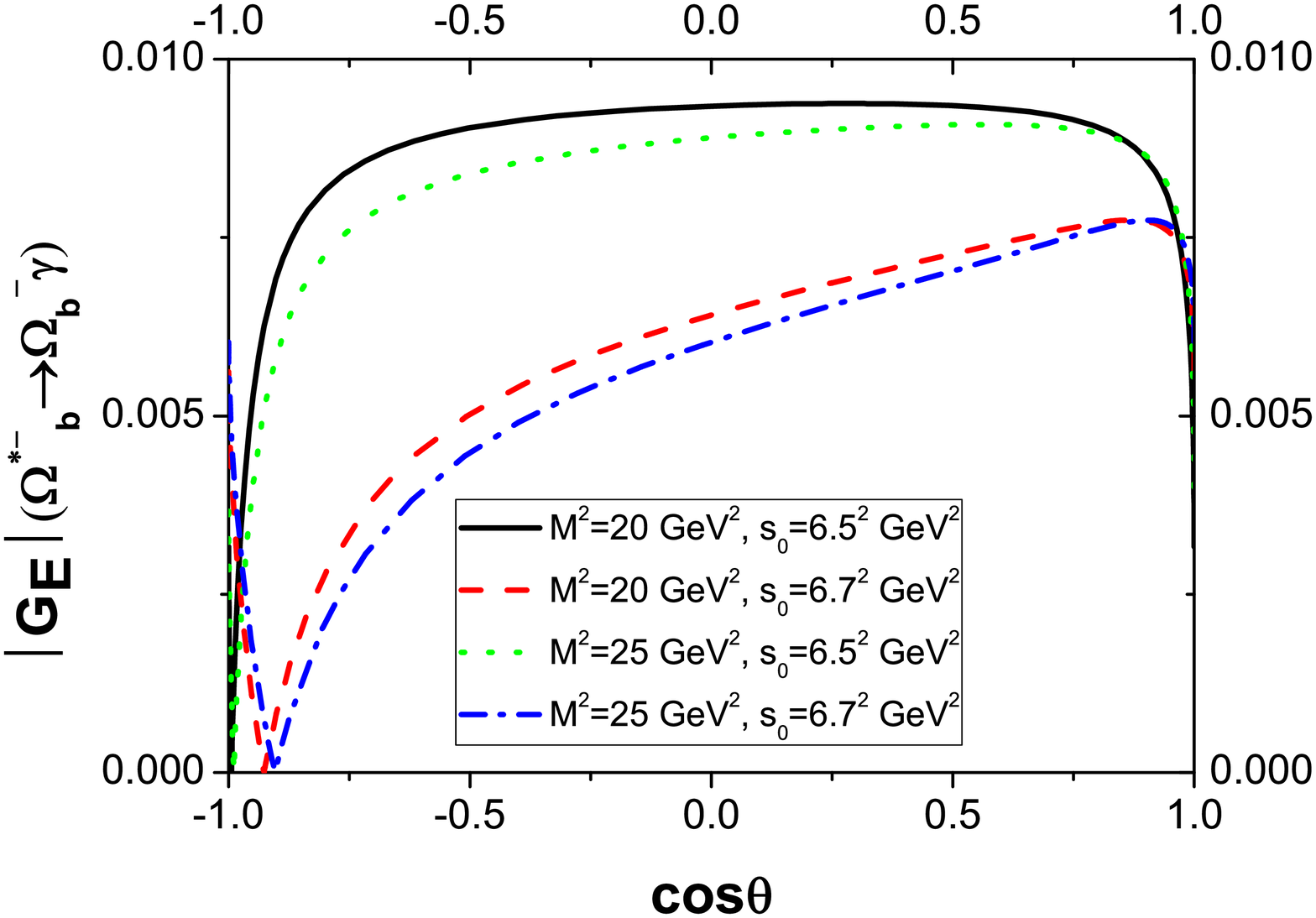}
\caption{\textbf{Left:} The dependence of the magnetic dipole moment  $G_{M}$ for $\Omega^{*-}_{b}\rightarrow
\Omega^{-}_{b}\gamma$ on $cos\theta$. \textbf{Right:} The
dependence of the electric quadrupole moment f  $G_{E}$ for
$\Omega^{*-}_{b}\rightarrow \Omega^{-}_{b}\gamma$ on $cos\theta$.
} \label{fig1a}
\end{figure}

\begin{figure}[h!]
\includegraphics[width=8cm]{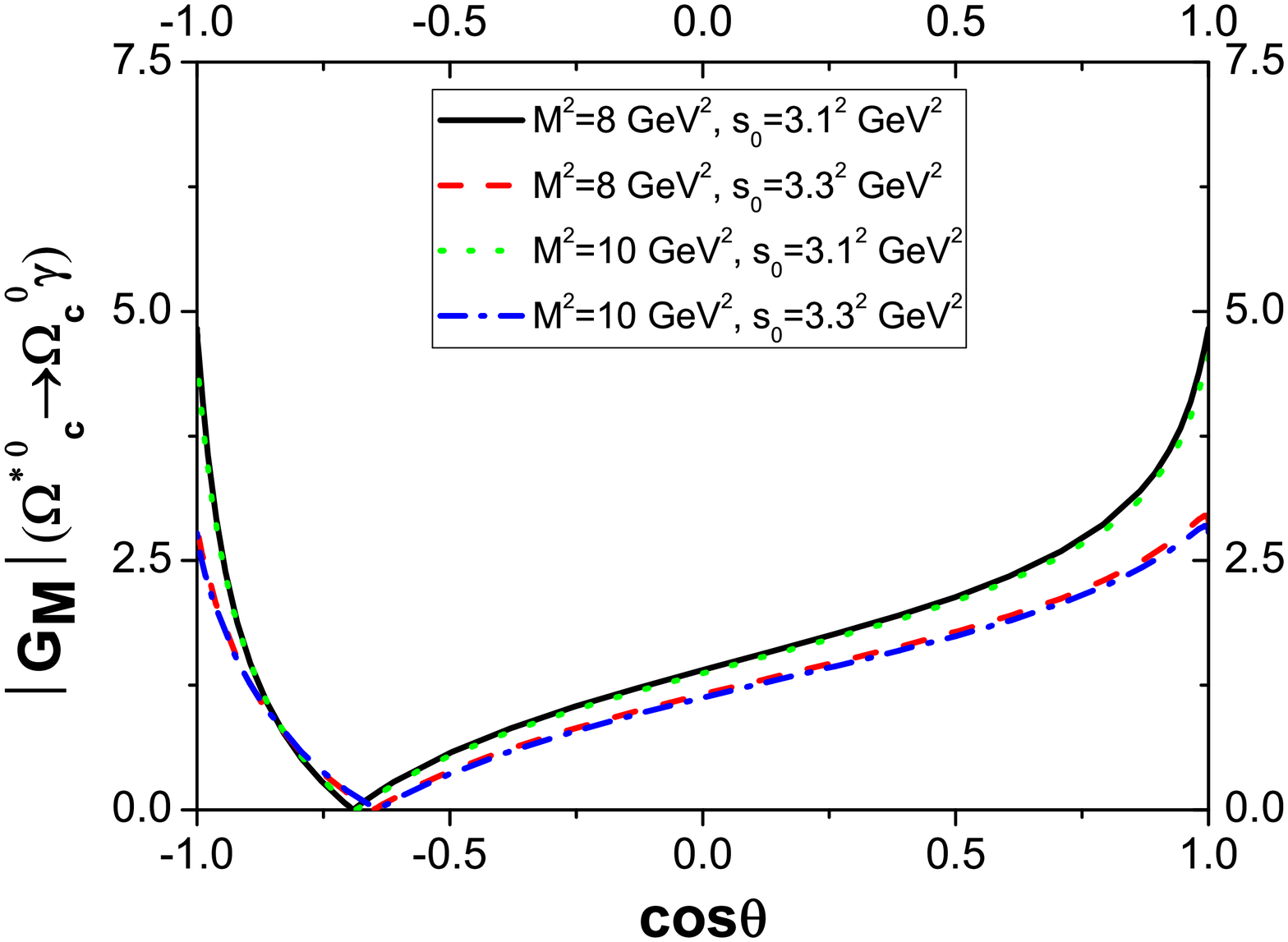}
\includegraphics[width=8cm]{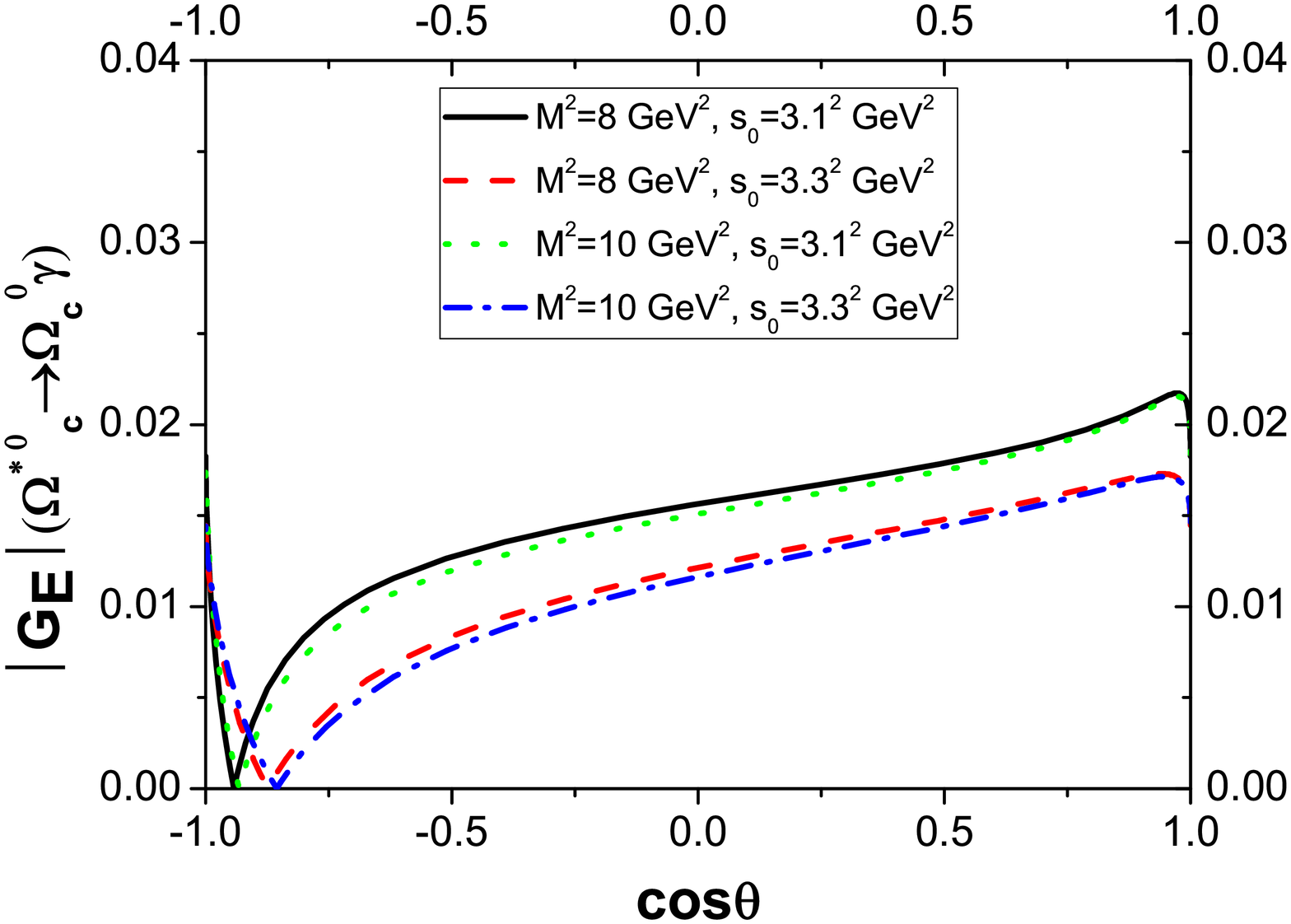}
\caption{ The same as Fig. \ref{fig1a}, but  for
$\Omega^{*0}_{c}\rightarrow \Omega^{0}_{c}\gamma$.} \label{fig2a}
\end{figure}

\begin{figure}[h!]
\includegraphics[width=8cm]{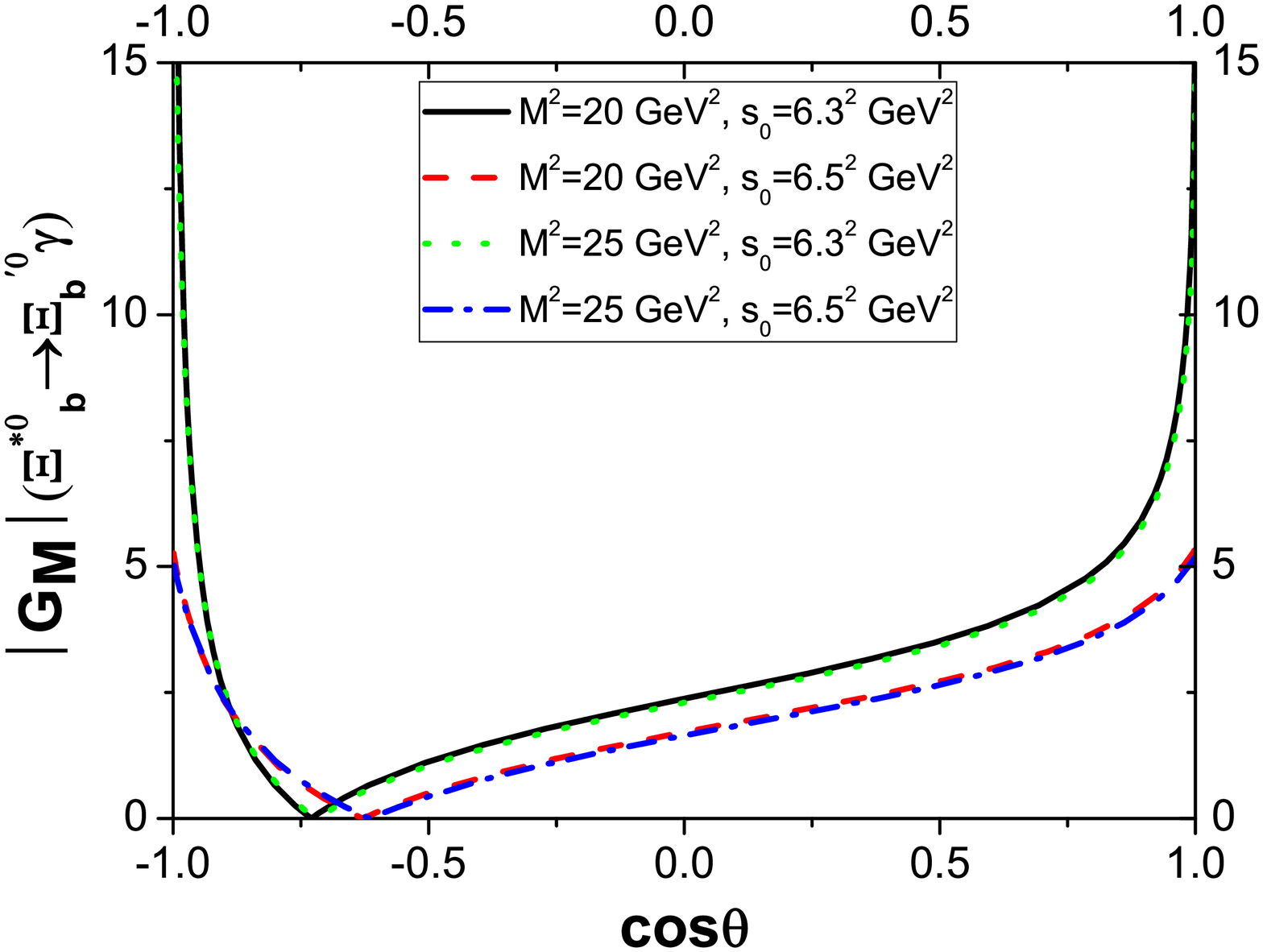}
\includegraphics[width=8cm]{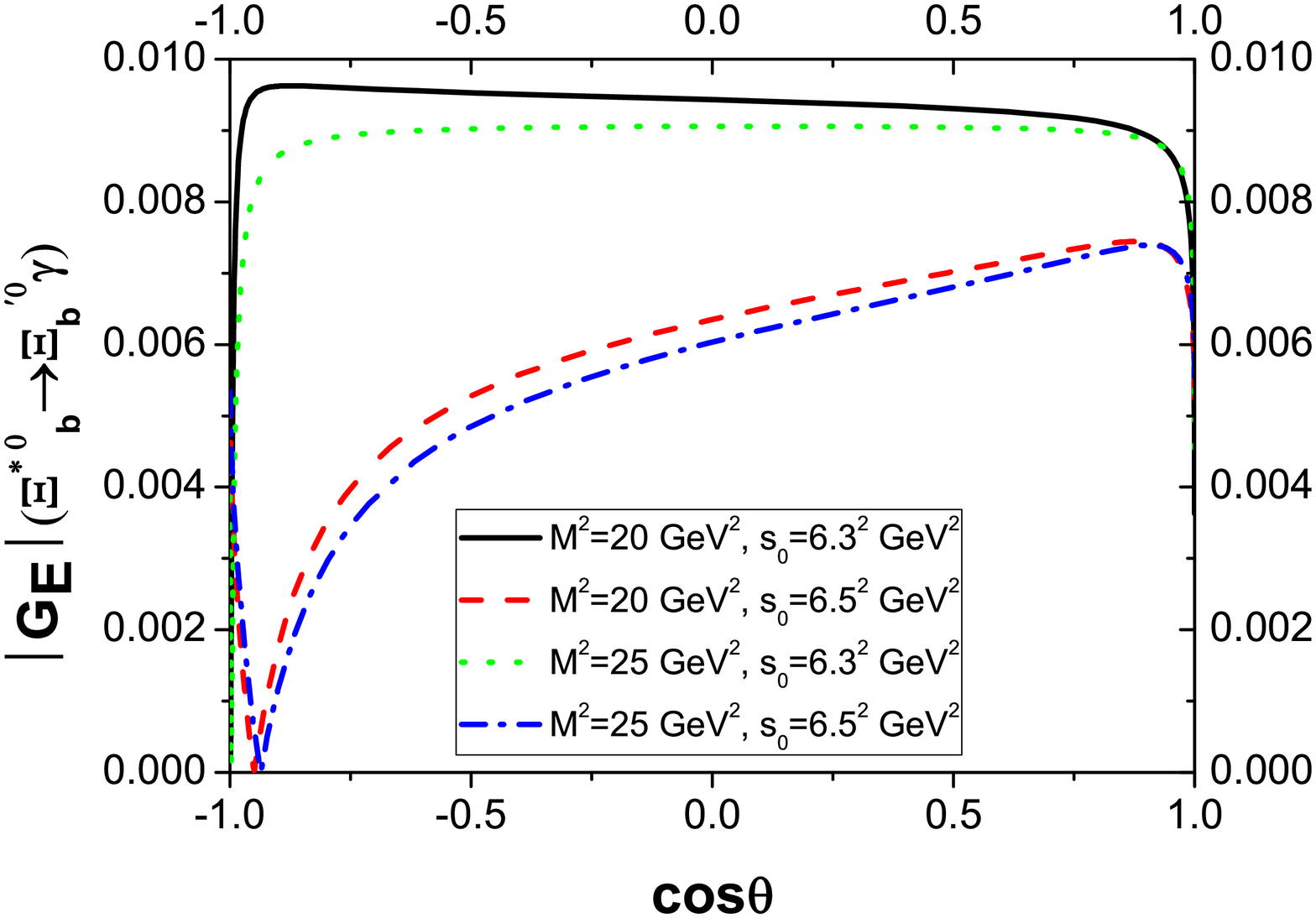}
\caption{The same as Fig. \ref{fig1a}, but  for
$\Xi^{*0}_{b}\rightarrow \Xi^{\prime 0}_{b}\gamma$.} \label{fig3a}
\end{figure}

\begin{figure}[h!]
\includegraphics[width=8cm]{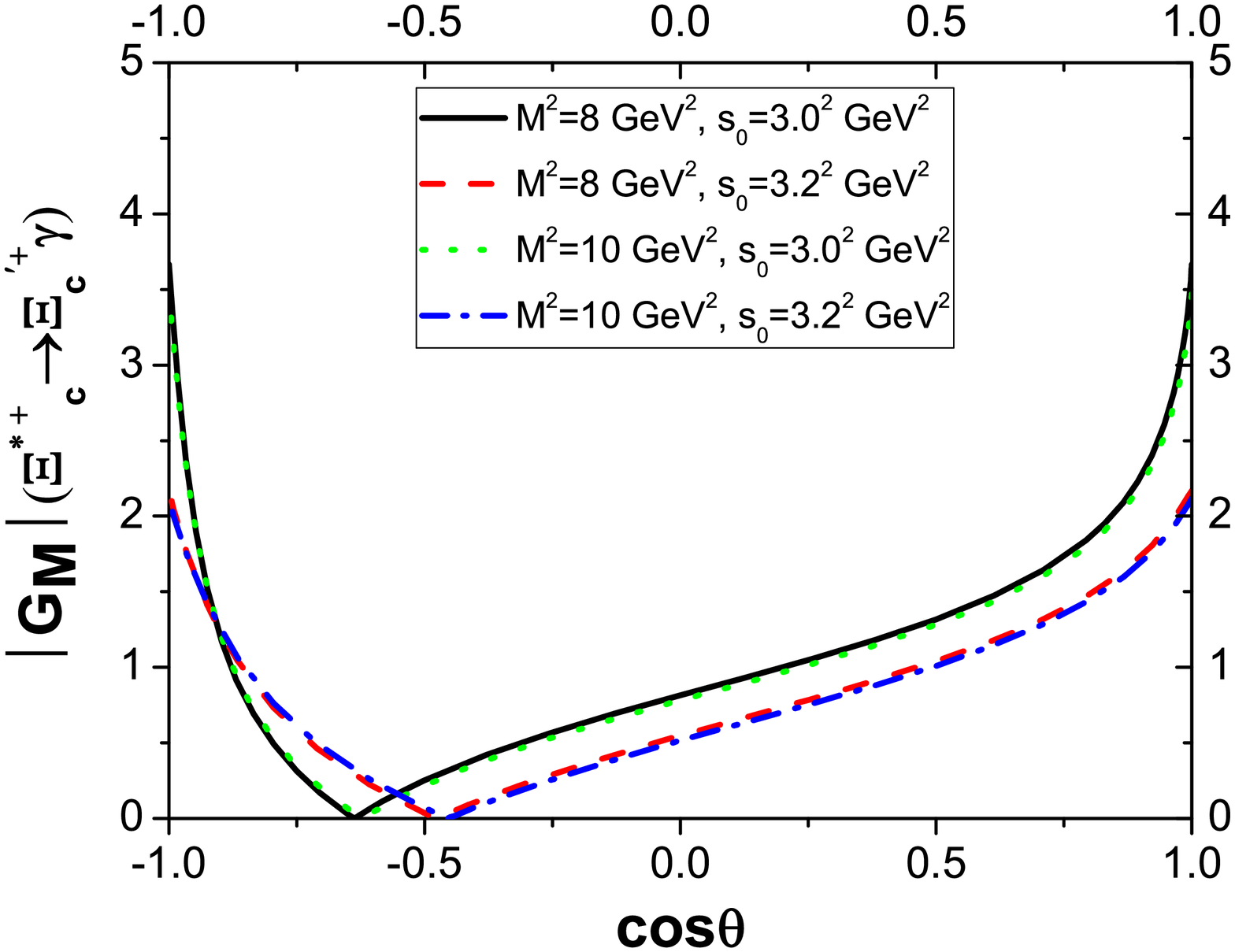}
\includegraphics[width=8cm]{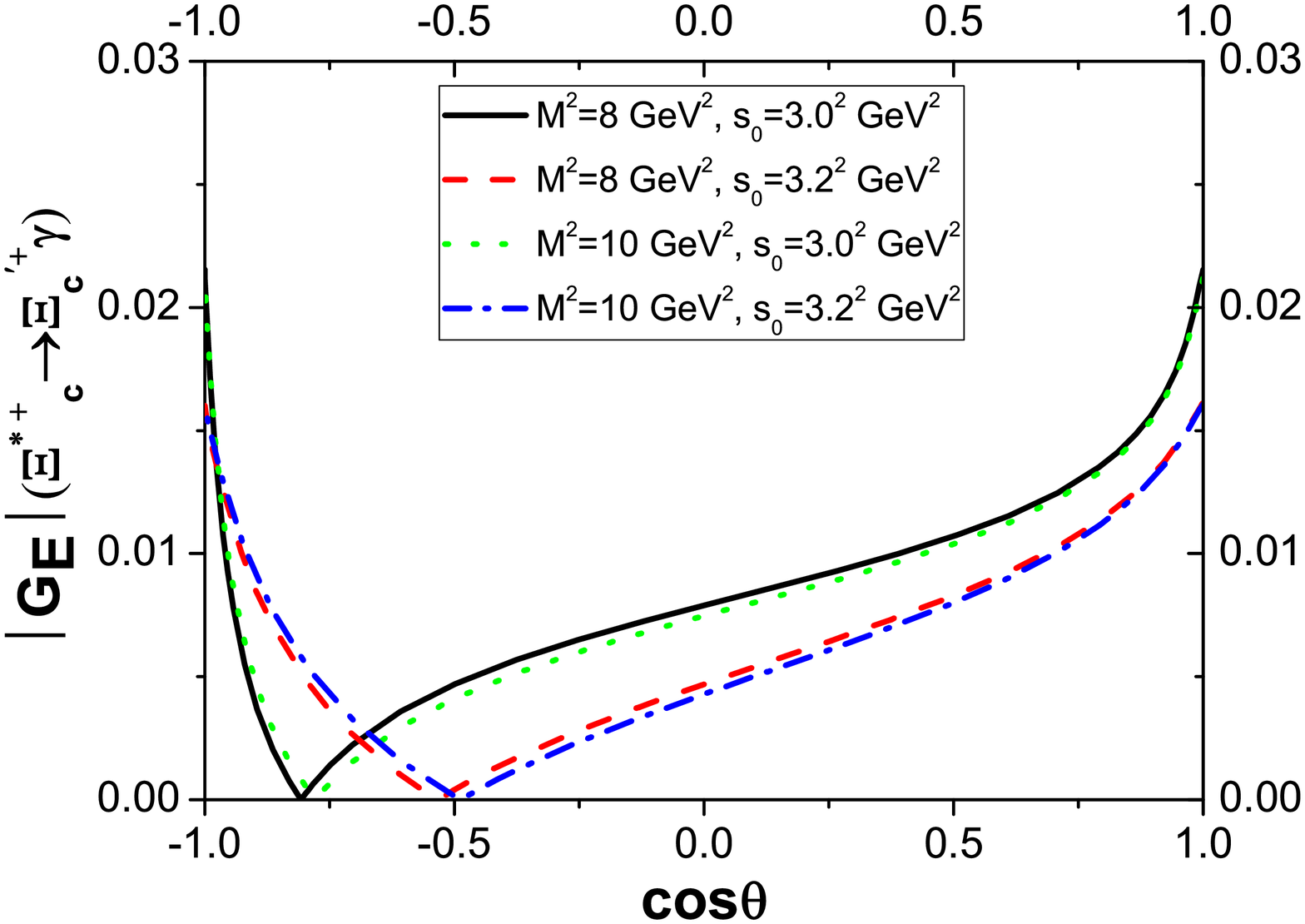}
\caption{The same as Fig. \ref{fig1a}, but  for
$\Xi^{*+}_{c}\rightarrow \Xi^{\prime +}_{c}\gamma$.} \label{fig4a}
\end{figure}

\begin{figure}[h!]
\includegraphics[width=8cm]{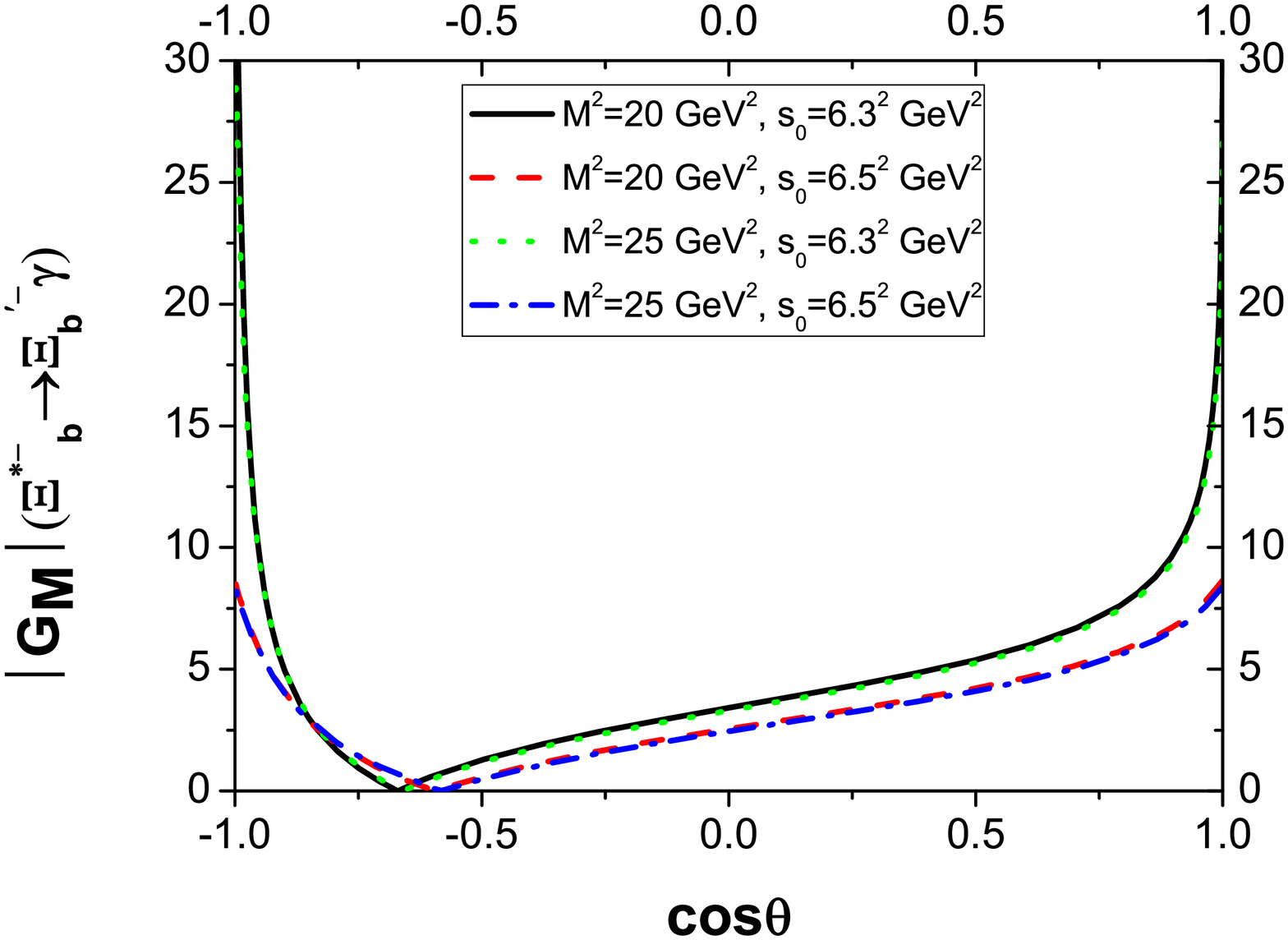}
\includegraphics[width=8cm]{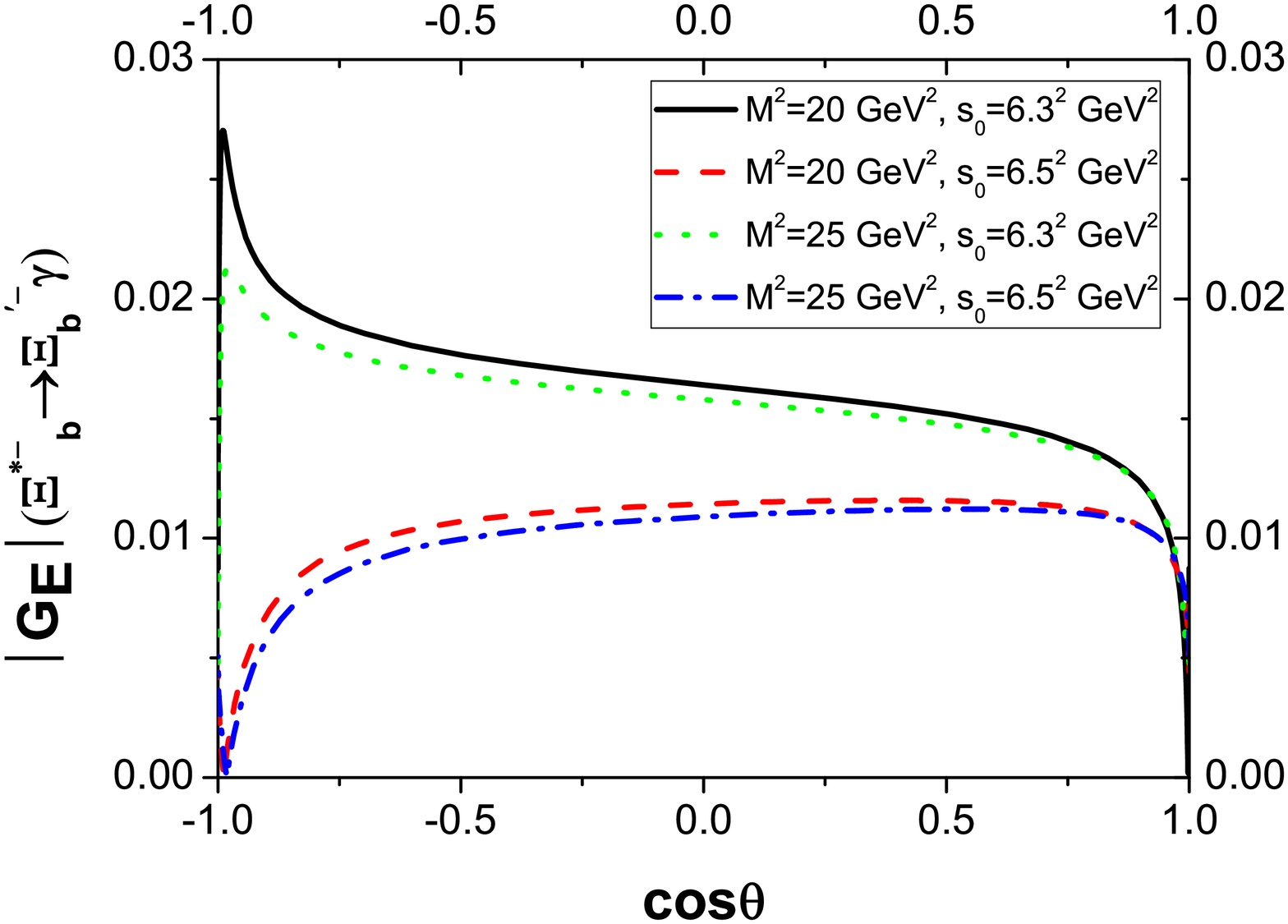}
\caption{The same as Fig. \ref{fig1a}, but  for
$\Xi^{*-}_{b}\rightarrow \Xi^{\prime -}_{b}\gamma$.} \label{fig5a}
\end{figure}

\begin{figure}[h!]
\includegraphics[width=8cm]{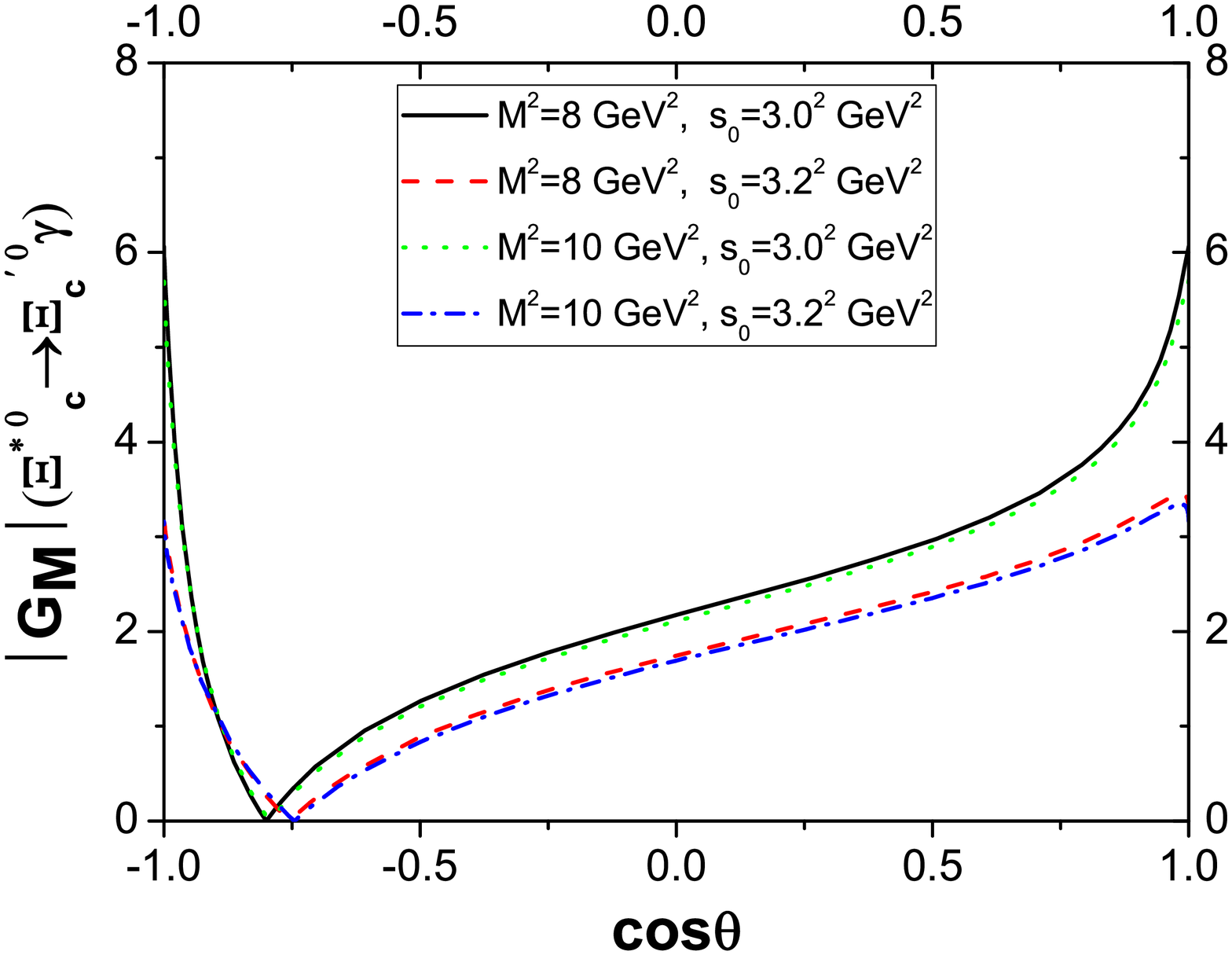}
\includegraphics[width=8cm]{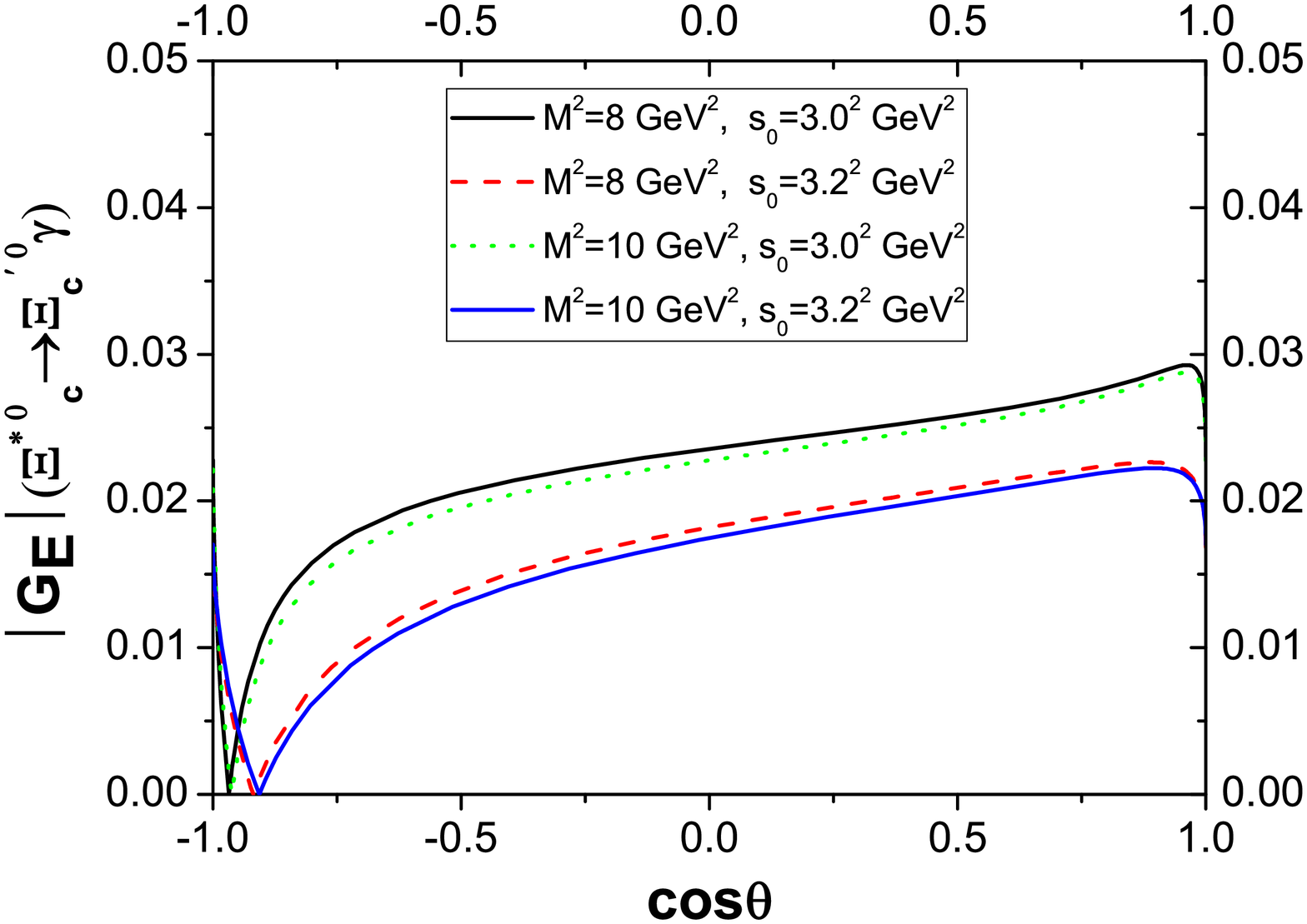}
\caption{The same as Fig. \ref{fig1a}, but  for
$\Xi^{*0}_{c}\rightarrow \Xi^{\prime 0}_{c}\gamma$.} \label{fig6a}
\end{figure}

 To determine the working regions for the general parameter $\beta$ at different radiative channels, we depict the dependence of the results on this parameter at different fixed values of the
Borel mass parameter and continuum threshold in figures \ref{fig1a}-\ref{fig6a}. Note that instead of $\beta$ we use $cos\theta$, where $\beta=tan\theta$. The interval $-1\leq cos\theta \leq1$
corresponds to $\beta$ between $-\infty$ to $+\infty$ which we shall consider in our calculations. The numerical results show that the values of $G_{E}$ are negligibly small and therefore we consider
  only dependence of
 $G_{M}$  on $\beta$, in order to find its working region.


From  figures \ref{fig1a}-\ref{fig6a}, we obtain the  region
$-0.25\leq\cos\theta\leq 0.5$ common for all radiative transitions under consideration, at which the dependence of the $G_{M}$  on $cos\theta$ is relatively weak.
In most of the figures related to the magnetic dipole moment, the Ioffe current which corresponds to
$cos\theta\simeq-0.71$ remains out of the reliable region.

Considering the working regions for the auxiliary parameters, photon DAs and other input parameters, we extract the values of the magnetic dipole moment $G_{M}$ and the electric quadrupole
  moment $G_{E}$ correspond to the considered  radiative transitions as presented in  table \ref{tabel3}. For comparison, we also present the predictions of VDM \cite{zamir} on $G_{M}$
and $G_{E}$ in this table. From this table we see that, considering the errors in our results, our predictions are comparable with those of VDM on the magnetic dipole moment $G_{M}$ for all transitions except that
$\Omega^{*-}_{b}\rightarrow \Omega^{-}_{b}\gamma$ which our result is considerably small compared to that of VDM. In both models, the values of $G_{E}$ are negligibly small for all considered channels.

\begin{table}[h]
\centering
\begin{tabular}{|c||c|c|c|c|}\hline
 & $|G_{M}|$ (PW) & $|G_{E}|$ (PW) &$|G_{M}|$ (VDM) \cite{zamir}&$|G_{E}|$ (VDM) \cite{zamir}\\\cline{1-5}
\hline\hline $\Omega^{*-}_{b}\rightarrow \Omega^{-}_{b}\gamma$
&$1.715\pm0.498$&$0.007$ &4.52 &0.034\\
$\Omega^{*0}_{c}\rightarrow
\Omega^{0}_{c}\gamma$&$1.337\pm0.374$ &$0.013$ & 2.17&0.026\\
 $\Xi^{*0}_{b}\rightarrow \Xi^{\prime 0}_{b}\gamma$&$2.003\pm0.601$&$0.006$&2.93 &0.017\\
 $\Xi^{*+}_{c}\rightarrow \Xi^{\prime +}_{c}\gamma$&$0.688\pm0.192$&$0.006$& 1.33&0.019\\
 $\Xi^{*-}_{b}\rightarrow \Xi^{\prime -}_{b}\gamma$&$3.037\pm0.881$&$0.011$&4.63 &0.021\\
$\Xi^{*0}_{c}\rightarrow \Xi^{\prime
0}_{c}\gamma$&$1.924\pm0.556$&$0.019$& 2.20&0.026\\\cline{1-5}
 \end{tabular}
 \vspace{0.8cm}
\caption{The   absolute values of the  magnetic dipole moment $|G_{M}|$
and electric quadrupole  moment $|G_{E}|$ for the corresponding
radiative decays  in units of  natural magneton. PW means present work and VDM refers to the vector dominance model.
}\label{tabel3}
\end{table}

At the end of this section we would like to present the decay width for the radiative transitions under consideration. Considering the transition matrix element in Eq. (4) and definitions of the magnetic dipole and electric
quadrupole moments in terms of form factors $G_1$ and $G_2$, we get the following formula for the widths of the corresponding transitions:
\begin{eqnarray}
\Gamma=3 \frac{\alpha}{32}\frac{(m_1^2-m_2^2)^3}{m_1^3m_2^2}(G_M^2+3G_E^2).
\end{eqnarray}
Using the numerical values for the magnetic dipole and electric quadrupole moments as well as the QCD sum rules predictions for the baryon masses, viz. $\Omega^{*}_{b}=(6.17\pm0.15)~GeV$,
$\Omega^{*}_{c}=(2.79\pm0.19)~GeV$, $\Xi^{*}_{b}=(6.02\pm0.17)~GeV$, $\Xi^{*}_{c}=(2.65\pm0.20)~GeV$, $\Omega_{b}=(6.11\pm0.16)~GeV$,
$\Omega_{c}=(2.70\pm0.20)~GeV$, $\Xi^{'}_{b}=(5.96\pm0.17)~GeV$ and $\Xi^{'}_{c}=(2.56\pm0.22)~GeV$ \cite{wang11,wang12}, we get the values for the widths as presented in table 4. 
For comparison, we also  depict the existing predictions form
the VDM in the same table. Looking at this table we see that our results are overall comparable in orders of magnitudes with the results of  \cite{zamir} except
 for $\Omega^{*-}_{b}\rightarrow \Omega^{-}_{b}\gamma$
channel at which our result is roughly one order of magnitude small compared to that of \cite{zamir}.  When we compare our results with those of
\cite{wang1,wang2}, we see considerable differences in orders of magnitudes between two models predictions  except for $\Omega^{*0}_{c}\rightarrow
\Omega^{0}_{c}\gamma$ and $\Xi^{*+}_{c}\rightarrow \Xi^{\prime +}_{c}\gamma$ channels that our predictions are in the same orders of magnitude with those of \cite{wang1,wang2}. 
The big differences among our results, \cite{zamir} and   \cite{wang1,wang2}
 may be attributed to the different baryon masses that are used  since
the width in Eq. (19) is very sensitive to the masses
of the initial and final baryons.

\begin{table}[h]
\centering
\begin{tabular}{|c||c|c|c|}\hline
 & $\Gamma$ (PW) & $\Gamma$ (VDM) \cite{zamir}&$\Gamma$ (VDM) \cite{wang1,wang2}\\\cline{1-4}
\hline\hline $\Omega^{*-}_{b}\rightarrow \Omega^{-}_{b}\gamma$
&$0.092$&2.873 &0.00074 \\ $\Omega^{*0}_{c}\rightarrow
\Omega^{0}_{c}\gamma$&$0.932$ & 1.439& 1.16\\
 $\Xi^{*0}_{b}\rightarrow \Xi^{\prime 0}_{b}\gamma$&$0.131$&0.281& 0.047\\
 $\Xi^{*+}_{c}\rightarrow \Xi^{\prime +}_{c}\gamma$&$0.274$&0.485&0.96 \\
 $\Xi^{*-}_{b}\rightarrow \Xi^{\prime -}_{b}\gamma$&$0.303$&0.702&0.066 \\
$\Xi^{*0}_{c}\rightarrow \Xi^{\prime 0}_{c}\gamma$&$2.142$&1.317&
0.12\\\cline{1-4}
 \end{tabular}
 \vspace{0.8cm}
\caption{Widths of the corresponding radiative transitions in KeV.}\label{tabel4}
\end{table}

In summary, we  have calculated the transition magnetic dipole moment $G_{M}$ and
electric quadrupole  moment $G_{E}$ as well as decay width for the radiative  $\Omega_{Q}^{*}\rightarrow\Omega_{Q}\gamma$ and
$\Xi_{Q}^{*}\rightarrow\Xi^{\prime}_{Q}\gamma$  transitions  within the light cone QCD
  sum rule approach and compared the results with the predictions of the VDM. Considering the recent progresses on the identification and spectroscopy of the heavy baryons, we hope  it will be possible to study these
radiative decay channels at the experiment in near future.

\section{Acknowledgment}
K. A. and H. S. would like to thank TUBITAK for their partial
financial support through the project  114F018.
%

%


\begin{thebibliography}{99}
%
\bibitem{Rbwtq02} M. Mattson et.al, SELEX Collaboration,  Phys. Rev. Lett. 89, 112001 (2002).
\bibitem{Rbwtq03} A. Ocherashvili et.al, SELEX Collaboration,  Phys.  Lett. B 628, 18 (2005).
\bibitem{zamir} T. M.  Aliev, M. Savci, V. S. Zamiralov, Mod. Phys. Lett. A 27, 1250054 (2012).
\bibitem{zamir2} T. M.  Aliev, K. Azizi, M. Savci, V. S. Zamiralov, Phys. Rev.D 83, 006007 (2011).
\bibitem{azizi} T. M. Aliev, K. Azizi, A. Ozpineci, Phys. Rev.D 79, 056005 (2009).
\bibitem{banuls} M. Banuls, A. Pich, I. Scimemi, Phys. Rev. D 61, 094009 (2000).
\bibitem{cheng}  H. Y. Cheng, et. al, Phys. Rev. D 47, 1030 (1993).
\bibitem{tawfig}  S. Tawfig, J. G. Koerner, P. J.O'Donnel, Phys. Rev. D 63, 034005 (2001).
\bibitem{ivanov}  M. A. Ivanov, et. al, Phys. Rev. D 60, 094002 (1999).
\bibitem{zhu}  S. L. Zhu, Y. B. Dai, Phys. Rev. D 59, 114015 (1999).

\bibitem{onalti} H. F. Joens, M. D. Scadron, Ann. Phys. 81, 1 (1973).
\bibitem{onyedi} R. C. E. Devenish, T.S. Eisenschitz and J. G. Korner, Phys. Rev. D  14, 3063 (1976).
\bibitem{Bagan} E. Bagan, M. Chabab, H. G. Dosch and S. Narison, Phys. Lett. B 278, 369 (1992).
\bibitem{Balitsky} I. I.  Balitsky,  V. M.  Braun,  Nucl. Phys.  B  311, 541 (1989).
\bibitem{Braun2} V. M.  Braun, I. E. Filyanov, Z. Phys. C 48, 239 (1990).
\bibitem{revised2} K. G. Chetyrkin, A. Khodjamirian, A. A. Pivovarov, Phys. Lett. B  661, 250 (2008);
 I. I. Balitsky, V. M. Braun, A. V. Kolesnichenko, Nucl. Phys. B   312, 509 (1989).
\bibitem{Ball} P.  Ball,  V. M.  Braun, N. Kivel,  Nucl. Phys.  B  649, 263 (2003).
\bibitem{Belyaev} V. M. Belyaev,  B. L.  Ioffe, JETP  56, 493 (1982).
\bibitem{Rohrwild} J. Rohrwild, JHEP  0709, 073 (2007).
\bibitem{balitskibal} I. I.  Balitsky, A. V. Kolesnichenko, A. V. Yung, Yad. Fiz. 41, 282 (1985).
\bibitem{Kogan} V. M. Belyaev,  I. I.  Kogan, Yad. Fiz.  40, 1035 (1984).
\bibitem{wang11}  Zhi-Gang Wang, Eur. Phys. J. C 68, 459 (2010).
\bibitem{wang12}  Zhi-Gang Wang, Phys. Lett. B 685, 59 (2010).
\bibitem{wang1}  Zhi-Gang Wang, Eur. Phys. J. A 44, 105 (2010).
\bibitem{wang2} Zhi-Gang Wang,  Phys. Rev. D 81, 036002 (2010).



\end{thebibliography}
\end{document}